\documentclass[a4paper,12pt]{article}
\pdfoutput=1

\usepackage{amsmath}
\usepackage{amssymb}
\usepackage[small]{caption}
\usepackage{cite}
\usepackage[margin=1in]{geometry}
\usepackage{enumerate}
\usepackage[multiple]{footmisc}
\usepackage{graphicx}
\usepackage{physics}
\usepackage{slashed}
\usepackage{textcomp}
\usepackage[nottoc]{tocbibind}
\usepackage{xcolor}

\definecolor{mediumblue}{rgb}{0,0,0.8}
\usepackage{hyperref}
\hypersetup{
  linktocpage=true,
  colorlinks=true,
  citecolor=mediumblue,
  filecolor=mediumblue,
  linkcolor=mediumblue,
  urlcolor=mediumblue,
}
\newcommand{\mailref}[1]{\href{mailto:#1}{#1}}

\graphicspath{{./}{./figures/}}
\numberwithin{equation}{section}
\allowdisplaybreaks%

\DeclareMathOperator*{\argmin}{argmin}

\def\thefootnote{\fnsymbol{footnote}}

\begin{document}

\begin{titlepage}
  \begin{flushright}
    \texttt{CTPU-PTC-21-31}
  \end{flushright}

  \bigskip

  \begin{center}
    \bf \Large
    Could \boldmath{$M_{T2}$} be a singularity variable?
  \end{center}

  \medskip

  \begin{center}
    Chan~Beom~Park\footnote{E-mail: \texttt{\mailref{cbpark@ibs.re.kr}}}
  \end{center}

  \begin{center}
    \it Center for Theoretical Physics of the Universe,
    Institute for Basic Science (IBS),\\
    55 Expo-ro, Yuseong-gu, Daejeon 34126, Korea\\[0.2cm]
  \end{center}

  \medskip

  \begin{abstract}
    The algebraic singularity method is a framework for analyzing
    collider events with missing energy.
    It provides a way to draw out a set of singularity variables that
    can catch singular features originating from the projection of
    full phase space onto the observable phase space of measured
    particle momenta.
    It is a promising approach applicable to various physics processes
    with missing energy but still requires more studies for use in
    practice.
    Meanwhile, in the double-sided decay topology with an invisible
    particle on each side, the $M_{T2}$ variable has been known to be
    a useful collider observable for measuring particle masses from
    missing energy events or setting signal regions of collider
    searches.
    We investigate the relation between the two different types of
    kinematic variables in double-sided decay topology. We find that the
    singularity variables contain the $M_{T2}$ variable in many cases,
    although the former is not a strict superset of the latter.
  \end{abstract}
\end{titlepage}

\renewcommand{\thefootnote}{\arabic{footnote}}
\setcounter{footnote}{0}

\setcounter{tocdepth}{2}
\noindent\rule{\textwidth}{0.3pt}\vspace{-0.4cm}\tableofcontents
\noindent\rule{\textwidth}{0.3pt}

\section{Introduction}

\noindent
The $M_{T2}$ variable was devised to measure supersymmetric particle
masses in collider events with missing energy due to invisible
particles in the final state~\cite{Lester:1999tx, Barr:2003rg}.
Although the discovery of supersymmetry remains to be made, the
$M_{T2}$ variable has served as the main observable in various physics
analyses at hadron colliders from the measurement of top quark
mass and properties~\cite{Cho:2008cu, CDF:2009zjw, Guadagnoli:2013xia,
  CMS:2017znf} to the searches for
supersymmetric particles~\cite{Barr:2009wu, ATLAS:2021hza,
  CMS:2021cox}.
It has also directly influenced the inventions of other kinematic
methods and variables such as the $M_{T2}$-assisted on-shell (MAOS)
method~\cite{Cho:2008tj}, $M_{CT2}$~\cite{Cho:2009ve}, and
$M_2$~\cite{Cho:2014naa}.
We will present a brief review of the $M_{T2}$ variable in
Sec.~\ref{sec:mt2}.

Considering that the invention of the $M_{T2}$ variable started from
the attempt of generalizing the transverse mass~\cite{Barger:1983wf,
  Smith:1983aa} to deal with two identical decay chains, the
algebraic singularity method has emerged from the realization that the
phase space of particle momenta is the solution space of the system of
kinematic constraints~\cite{Kim:2009si}.
At collider experiments, we detect only visible particle momenta while
missing invisible ones. It can be regarded as the projection of the
full phase space of collider event onto the visible subspace.
The algebraic singularity method proposes that one can construct
singularity variables, which unveil singularities caused by the
projection of phase space, by exploiting all the available kinematic
constraints in a specific way.
Compared to the $M_{T2}$ variable, the algebraic singularity method is
a general framework applicable to any decay topology of physics
processes containing invisible particles in the final state.
The formal description of the algebraic singularity method and the
singularity variables are given in Sec.~\ref{sec:singularity_method}.

The author of Ref.~\cite{Kim:2009si} mentioned that the algebraic
singularity method was initiated due to his desire to understand the
$M_{T2}$ variable, and he could eventually derive the $M_{T2}$
variable by using the algebraic singularity method~\cite{ref:kim}.
However, in Ref.~\cite{Kim:2009si}, the relation between the $M_{T2}$
variable and the algebraic singularity method was not elucidated,
and it has never been considered in the following
works by different authors~\cite{Rujula:2011qn, DeRujula:2012ns,
  Matchev:2019bon, Park:2020rol}.
Moreover, in Ref.~\cite{Kim:2021pcz}, it has even been argued that
$M_{T2}$ is not associated with the singularity variables.
On the other hand, in other studies, it has been realized that $M_{T2}$
is not an ad hoc variable, but it sets a boundary of the mass
region consistent with the kinematic constraints, which are indeed the
defining polynomials of phase space that are used in the algebraic
singularity method~\cite{Cheng:2008hk, Barr:2009jv}.
These seemingly discrepant arguments and claims have motivated us to
examine the relation between the $M_{T2}$ variable and the
corresponding singularity variables.
If they turn out to be truly orthogonal, the algebraic singularity
method will provide new kinematic variables that can be employed in
the physics analyses where either or both of the $M_{T2}$ variable and
the new variables are applicable.
Otherwise, if they are related to each other in any manner, it may
still lead to a deeper understanding of the methods.
Either way, the algebraic singularity method is an interesting
framework for studying various decay topologies with missing energy.
We investigate the relation between the $M_{T2}$ and the singularity
variable in Sec.~\ref{sec:comp}.
Then, the final section is devoted to conclusions.

Before embarking on a discussion of the above issues, a few
comments are in order on the applicability of the algebraic
singularity method to physics analyses at colliders.
As mentioned above, the algebraic singularity method is, in principle,
a framework for analyzing the collider events of any decay topology
with missing energy rather than a collider variable designed for
specific physics processes.
The method is still widely unknown and largely unexplored in physics
community due to its use of abstract mathematics and the lack of
concrete prescriptions with practical examples.
A set of worked-out examples of deriving the singularity variables
using the method has appeared only very
recently~\cite{Matchev:2019bon}.
One way to appreciate the potential of the method would be to compare
the singularity variables with the known collider variables.
In a single two-body decay with an invisible particle in the final
state, it has been shown that the well-known transverse
mass~\cite{Barger:1983wf, Smith:1983aa} can be
derived by using the method~\cite{Rujula:2011qn, DeRujula:2012ns,
  Matchev:2019bon, NewPhysicsWorkingGroup:2010hhy}.
We will show the derivation of the transverse mass in
Sec.~\ref{sec:singularity_method} by using the Gr\"obner basis as
proposed in the original literature~\cite{Kim:2009si} to help the
reader understand the method.
It has also been found that the $\Delta_4$
observable~\cite{Agrawal:2013uka, Debnath:2018azt}, appearing as a
factor in four-body phase space~\cite{Byers:1964ryc}, corresponds to
the singularity variable~\cite{Matchev:2019bon}.
We anticipate that our attempt of comparing the $M_{T2}$ with the
singularity variables in double-sided decay topology will also help
the reader to gain an understanding of the method and to apply the
method to find the signals of unexplored decay topologies at colliders.

Though we will not pursue here, there is room for improvement
in the way of identifying the singularities of phase space.
As noted above, the algebraic singularity method is to capture the
kinematic singularities caused by the projection of the full phase
space onto the visible subspace.
In the original literature, it has been argued that the Gr\"obner
basis of kinematic constraint equations would be particularly useful
for identifying singular points on the visible subspace.
Meanwhile, the other following works~\cite{Rujula:2011qn,
  DeRujula:2012ns, Matchev:2019bon, Park:2020rol} have obtained the
singularity variables without employing Gr\"obner bases.
As will be seen in Sec.~\ref{sec:singularity_method}, Gr\"obner bases
have a useful property for identifying singularities.
However, the expressions appear to be much more complex than the
original system of kinematic constraints, and computing Gr\"obner
bases can be challenging, depending on the form and the numbers of the
polynomial equations.
We may attempt a different formulation of the method by systematically
transforming the system of kinematic constraints into a more refined
space of polynomial equations that one can more handily extract
singularities.
Another possible direction worth exploring is to combine the method
with machine learning techniques.
The singularity variables can be used as input features for training
collider data, and it is conceivable to devise an architecture
encoding the algorithm for finding the singularities.
We stress that there is much more to explore in further developments
and finding the applications of the algebraic singularity method to
collider data analyses.

\section{\label{sec:mt2}The \boldmath{$M_{T2}$} variable}

\noindent
We begin our discussion with the definition and the properties of the
$M_{T2}$ variable before looking into the algebraic singularity
method.
The $M_{T2}$ variable was devised as a generalization of the
transverse mass for measuring the masses of supersymmetric particles
produced in a pair at hadron colliders,
and hence it is often called the {\em stransverse mass} in literature.
Supersymmetric processes yield missing energy events if the
lightest supersymmetric particle is electrically neutral and stable on
a time scale well beyond the scale of the detector.
The processes typically have the decay topology of
\begin{equation}
  Y + \bar Y + U \longrightarrow v_1(p_1) \chi(k_1) + v_2(p_2)
  \bar \chi(k_2) + U(u),
  \label{eq:topology}
\end{equation}
where $v_1$ and $v_2$ are visible objects such as charged leptons and
jets, and $\chi$ is the invisible particle corresponding to the
culprit responsible for the missing energy.
For example, consider the process of pair-produced gluinos, each of
which decays into the final state of two quarks plus the lightest
neutralino, $\tilde g \tilde g\to q \bar q \tilde\chi_1^0
+ q^\prime \bar q^\prime \tilde\chi_1^0$.
In this process, $v_1$ and $v_2$ correspond to the two-quark-jet
systems, and $\chi$ is the invisible neutralino.
Hereafter, the index $a = 1$, $2$ denotes each side of the decays
in~\eqref{eq:topology}.
$U$ is the upstream-momentum object, which is not associated with the
decays of $Y$ and $\bar Y$, such as initial state radiation.
We will take into account the effect of upstream momentum in
Sec.~\ref{sec:comp}.
In this article, we assume that the decay topology is symmetric, i.e.,
$M_{Y} = M_{\bar Y}$ and $M_{\chi} = M_{\bar \chi}$.

For the double-sided decay topology~\eqref{eq:topology} and
the hypothesized invisible particle mass $M_\chi$,
the $M_{T2}$ variable is defined as
\begin{align}
  M_{T2} \equiv
  &\min_{\vb*{k}_{1T}, \, \vb*{k}_{2T} \in \mathbb{R}^2}
    \Big[ \max \Big\{
    M_{1T}(p_{1T}, \, k_{1T}, \, M_\chi) , \,
    M_{2T}(p_{2T}, \, k_{2T}, \, M_\chi) \Big\}\Big]
    \nonumber\\
  & \text{subject to}\,\,\vb*{k}_{1T} + \vb*{k}_{2T} =
    \slashed{\vb*{P}}_T ,
    \label{eq:MT2}
\end{align}
where $M_{aT}$ are transverse masses given by
\begin{equation}
  M_{aT}(p_{aT}, \, k_{aT}, \, M_\chi) = \Big[ m_a^2 +
  M_\chi^2 + 2 \Big( E_{aT} e_{aT} - \vb*{p}_{aT} \cdot \vb*{k}_{aT}
  \Big) \Big]^{1/2}
  \label{eq:MT}
\end{equation}
for the final-state particle momenta projected onto the
$(1+2)$-dimensional space:
\begin{equation}
  p_{aT} = \big(E_{aT}, \, \vb*{p}_{aT} \big) ,\quad
  k_{aT} = \big(e_{aT}, \, \vb*{k}_{aT} \big) .
\end{equation}
Here, $E_{aT} = (m_a^2 + \norm{\vb*{p}_{aT}}^2)^{1/2}$ and $e_{aT} =
(M_\chi^2 + \norm{\vb*{k}_{aT}}^2)^{1/2}$ are transverse energies, and
$m_a^2 = p_a^2$.
As will be seen shortly,
$\vb*{k}_{aT}$ in Eq.~\eqref{eq:MT} are not the true invisible
momenta, nor inputs inserted by hand, but are determined by the
minimization of the objective function of $M_{T2}$ under the
constraint on the missing transverse momentum,
\begin{equation}
  \slashed{\vb*{P}}_T = - \vb*{p}_{1T} - \vb*{p}_{2T} - \vb*{u}_T .
\end{equation}
Therefore, the $M_{T2}$ variable is a function of visible particle
momenta $p_a$, missing transverse momentum $\slashed{\vb*{P}}_T$, and
the guessed value of invisible particle mass $M_\chi$.
What makes the $M_{T2}$ variable useful for mass measurement is that
if one inserts a correct value of $M_\chi$, its distribution has an
endpoint at the parent particle mass, $M_Y$:
\begin{equation}
  M_{T2}(M_\chi = M_\chi^\text{true}) \leq M_Y.
  \label{eq:MT2_endpoint}
\end{equation}
Because the $M_{T2}$ variable is constructed by only kinematic
quantities, it applies to any process, not only the supersymmetric
ones, as long as the decay topology given in~\eqref{eq:topology} can
describe the process of interest.

The calculation of the $M_{T2}$ variable corresponds to the
{\em constrained} minimization problem of the objective function,
\begin{equation}
  f(\vb*{k}_{aT}) = \max \Big\{ M_{1T}(k_{1T}), \, M_{2T}(k_{2T})
  \Big\},
  \label{eq:MT2_objective}
\end{equation}
subject to the equality constraint given by
\begin{equation}
  \vb*{c}(\vb*{k}_a) = \vb*{k}_{1T} + \vb*{k}_{2T} -
  \slashed{\vb*{P}}_T = \vb*{0} .
  \label{eq:MT2_constraint}
\end{equation}
The problem can be made simpler by eliminating $\vb*{k}_{2T}$ using
the constraint~\eqref{eq:MT2_constraint}, $\vb*{k}_{2T} =
\slashed{\vb*{P}}_T - \vb*{k}_{1T}$. Then, the objective function
becomes
\begin{equation}
  f(\vb*{k}_{1T}) = \max \Big\{ M_{1T}(e_{1T}, \, \vb*{k}_{1T}), \,
  M_{2T}(e_{2T}, \, \slashed{\vb*{P}}_T - \vb*{k}_{1T}) \Big\},
\end{equation}
where $e_{2T} = (M_\chi^2 + \norm{\slashed{\vb*{P}}_T -
  \vb*{k}_{1T}}^2)^{1/2}$.
The calculation of $M_{T2}$ has now become
an {\em unconstrained} minimization on a function of two
variables, $\vb*{k}_{1T} = (k_{1x}$, $k_{1y})$.
The value of $\vb*{k}_{1T}$ at the minimum of the objective function
$f(\vb*{k}_{1T})$ is the $M_{T2}$ solution to the invisible particle
momenta,
\begin{equation}
  \widetilde{\vb*{k}}_{1T} = \argmin_{\vb*{k}_{1T} \in \mathbb{R}^2}
  f(\vb*{k}_{1T}) .
\end{equation}
Here, $\argmin$ stands for the argument of minimum, i.e., the point
$\vb*{k}_{1T}$ for which $f(\vb*{k}_{1T})$ attains its minimum.
It has been found that $\widetilde{\vb*{k}}_{1T}$ can provide a good
approximation to the invisible particle momenta along with on-shell
mass constraints~\cite{Cho:2008tj}.
Even though we have converted to a simpler problem of
unconstrained minimization over the fewer degrees of freedom, one
should employ a numerical algorithm, such as the quasi-Newton method,
to obtain the value of $M_{T2}$.
We refer the reader to Refs.~\cite{Cho:2015laa, Park:2020bsu} for a
review of the $M_{T2}$ variable in terms of minimization problems and
the numerical algorithms feasible for solving the problem.

\begin{figure}[tb!]
  \begin{center}
    \includegraphics[height=0.46\textwidth]{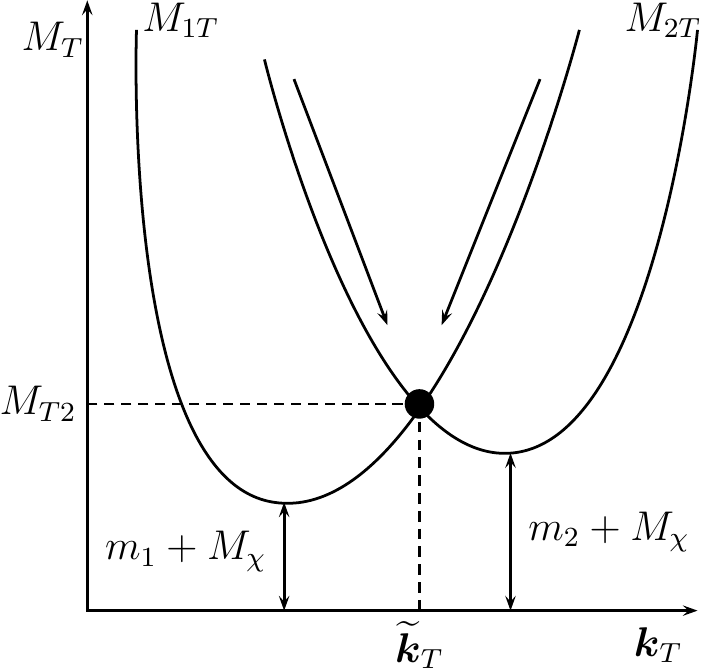}
    \includegraphics[height=0.46\textwidth]{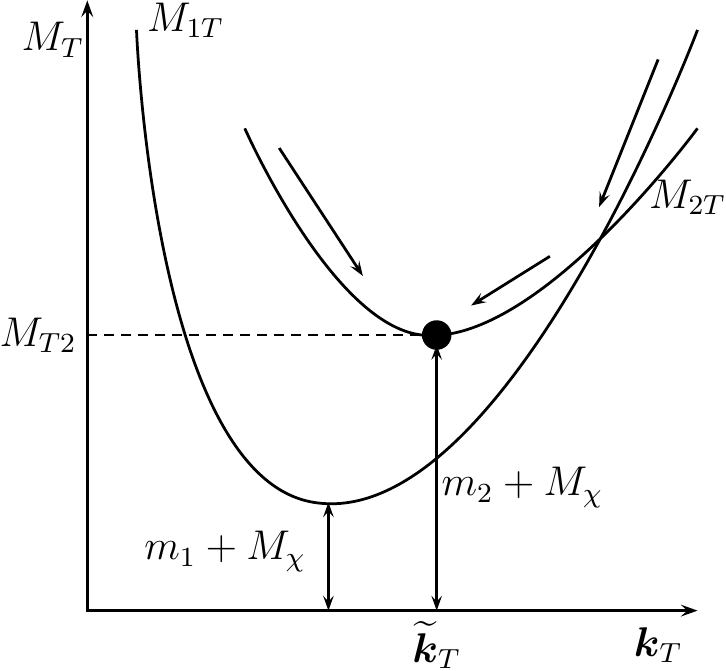}
  \end{center}
  \caption{\label{fig:mt2_conf}
    Schematic pictures of $M_{T2}$ for the balanced (left) and the
    unbalanced (right) configurations of transverse masses. The black
    dot in each plot corresponds to the $M_{T2}$ value obtained by
    minimization over the invisible particle momenta $\vb*{k}_T$, and
    $\widetilde{\vb*{k}}_T$ is the solution (or the minimizer) of
    $M_{T2}$.}
\end{figure}

Although we use numerical algorithms in practical physics analyses, it
is worth inspecting the analytic property of the $M_{T2}$ variable.
Because the objective function of $M_{T2}$ is a convex
function over the invisible particle momenta
$\vb*{k}_{aT}$~\cite{Park:2020bsu, Lim:2016ymd}, it has a unique local
minimum, which is automatically a global minimum.
Furthermore, as the transverse masses in Eq.~\eqref{eq:MT} are also
convex functions,
the $M_{T2}$ value is determined by the configuration of the
two convex functions, as depicted in Fig.~\ref{fig:mt2_conf}.
The stationary points of the transverse masses can be obtained by
requiring their gradients,
\begin{align}
  \frac{\partial M_{1T}^2}{\partial \vb*{k}_{1T}}
  &= 2 \left[
    \frac{E_{1T}}{e_{1T}} \vb*{k}_{1T} - \vb*{p}_{1T} \right] ,
  \nonumber\\
  \frac{\partial M_{2T}^2}{\partial \vb*{k}_{1T}}
  &= 2 \left[ -
    \frac{E_{2T}}{e_{2T}} \left( \slashed{\vb*{P}}_T - \vb*{k}_{1T}
    \right) + \vb*{p}_{2T} \right] ,
    \label{eq:MT_gradient}
\end{align}
to be vanishing.
Here, we have eliminated $\vb*{k}_{2T}$ by using the constraint on missing
transverse momentum given in Eq.~\eqref{eq:MT2_constraint}.
At each stationary point, we have the global minimum
of corresponding transverse mass as follows:
\begin{align}
  M_{1T} &= m_1 + M_\chi \quad \text{when } \vb*{k}_{1T} =
           \frac{e_{1T}}{E_{1T}} \vb*{p}_{1T} , \nonumber\\
  M_{2T} &= m_2 + M_\chi \quad \text{when } \vb*{k}_{1T} =
           \slashed{\vb*{P}}_T - \frac{e_{2T}}{E_{2T}} \vb*{p}_{2T} .
\end{align}
The configuration of the transverse masses are determined by
the visible particle momenta and missing transverse momentum.
In the configuration depicted in the left panel of
Fig.~\ref{fig:mt2_conf}, the $M_{T2}$ value is given by
\begin{equation}
  M_{T2} = M_{1T} = M_{2T}
  \label{eq:MT2_balanced}
\end{equation}
if both conditions below are satisfied:
\begin{align}
  M_{2T} &\geq m_1 + M_\chi \quad \text{when } \vb*{k}_{1T} =
           \frac{e_{1T}}{E_{1T}} \vb*{p}_{1T} , \nonumber\\
  M_{1T} &\geq m_2 + M_\chi \quad \text{when } \vb*{k}_{1T} =
           \slashed{\vb*{P}}_T - \frac{e_{2T}}{E_{2T}} \vb*{p}_{2T} .
           \label{eq:mt2_bal_cond}
\end{align}
This is called the {\em balanced} configuration.
Otherwise, if any of the conditions is violated, the $M_{T2}$ value
is given by the larger of the two global minima of the transverse
masses,
\begin{equation}
  M_{T2} = \max \Big[ m_1 + M_\chi , \, m_2 + M_\chi \Big] .
  \label{eq:MT2_unbal}
\end{equation}
See the right panel of Fig.~\ref{fig:mt2_conf}.
In this case, Eq.~\eqref{eq:MT2_balanced} does not hold, and it is
termed the {\em unbalanced} configuration.
The balanced and the unbalanced configurations span all possible cases
of the $M_{T2}$ solution: an event can be in either a balanced or an
unbalanced configuration.

The general analytic expression of $M_{T2}$ in balanced
configurations is hitherto unknown except for some special
cases~\cite{Lester:2007fq, Cho:2007dh, Cho:2009wh, Lester:2011nj,
  Lally:2012uj},
whereas the expression of the $M_{T2}$ variable in the case of the
unbalanced configuration is given by
Eq.~\eqref{eq:MT2_unbal}.\footnote{
  On the other hand,
  the analytic expressions for the endpoint of the $M_{T2}$
  distribution are known~\cite{Cho:2007qv, Barr:2007hy, Cho:2007dh,
    Burns:2008va, Konar:2009qr, Mahbubani:2012kx}.
}
When the upstream momentum in the decay
topology~\eqref{eq:topology} is vanishing, the expression of $M_{T2}$
in balanced configurations has been identified, and it is given as
follows~\cite{Lester:2007fq, Cho:2007dh}:
\begin{equation}
  M_{T2}^2 = M_\chi^2 + A_T + \sqrt{\left( 1 + \frac{4 M_\chi^2}{2 A_T -
        m_1^2 - m_2^2} \right) \left( A_T^2 - m_1^2 m_2^2 \right)} ,
  \label{eq:mt2_bal}
\end{equation}
where
\begin{equation}
  A_T = E_{1T} E_{2T} + \vb*{p}_{1T} \cdot \vb*{p}_{2T} .
\end{equation}
Likewise, the analytic expression of the $M_{T2}$ solution,
$\widetilde{\vb*{k}}_{1T}$, applicable in all respects, remains to be
identified.
In the case of vanishing upstream momentum, the derivation of the
$M_{T2}$ solution is capable but quite involved~\cite{ref:mt2solution}.
Nonetheless, in the plain case where the visible particles are
massless, $m_1 = m_2 = 0$, we can use a simple ansatz for the form of
the $M_{T2}$ solution in the balanced configuration, given as follows:
\begin{equation}
  \widetilde{\vb*{k}}_{1T} = \frac{1}{2} \Big[ \left( \alpha_T - 1 \right)
  \vb*{p}_{1T} - \left( \alpha_T + 1 \right) \vb*{p}_{2T} \Big] ,
  \label{eq:mt2_ksol}
\end{equation}
where $\alpha_T$ is a function of the visible particle momenta and
the invisible particle mass $M_\chi$.
By substituting the above expression into
\begin{equation}
  M_{T2}^2 = \frac{M_{1T}^2 + M_{2T}^2}{2}
\end{equation}
and using the expression of $M_{T2}$ in Eq.~\eqref{eq:mt2_bal} with
$m_1 = m_2 = 0$, we get
\begin{equation}
  \alpha_T = \sqrt{1 + \frac{2 M_\chi^2}{A_T}} .
\end{equation}
Here, $A_T = \norm{\vb*{p}_{1T}} \norm{\vb*{p}_{2T}} + \vb*{p}_{1T}
\cdot \vb*{p}_{2T}$.
If the invisible particle is massless as well, $M_\chi = 0$,
the expression~\eqref{eq:mt2_ksol} further simplifies as
\begin{equation}
  \widetilde{\vb*{k}}_{1T} = -\vb*{p}_{2T} .\label{eq:mt2_ksol_massless}
\end{equation}
Because we here assume that the upstream momentum is vanishing, the
missing transverse momentum is given by the negative vector sum of the
visible particle momenta from the decays of $Y$ and $\bar Y$,
$\slashed{\vb*{P}}_T = - \vb*{p}_{1T} - \vb*{p}_{2T}$. Therefore,
the invisible transverse momentum of the second decay chain is
given by
\begin{equation}
  \widetilde{\vb*{k}}_{2T} = -\vb*{p}_{1T} .
\end{equation}
This expression matches that identified in the study of $M_{T2}$ for
the $h \to W W \to \ell^+ \nu \ell^- \bar\nu$ process, where the
visible and invisible particles in the final state are
all massless~\cite{Choi:2009hn, Choi:2010dw}.
We will revisit the fully massless case in the last section before
conclusions.

\section{\label{sec:singularity_method}Algebraic singularity method}



\noindent
In this section, we describe the algebraic singularity method and then
turn to the discussion of the singularity variable in the double-sided
decay topology~\eqref{eq:topology}.
The algebraic singularity method was initiated from the realization
that the phase space of final-state particle momenta is an affine
variety in mathematics~\cite{Kim:2009si}.
Let $g_i$ be the polynomials of kinematic constraints such as
energy-momentum conservation and on-shell mass relations.
The total degrees of the polynomials $g_i$ are at most two.
The final-state particle momenta $P_j$ reside in the set of all
solutions of the polynomial equations,
\begin{equation}
  \Pi(g_1, \, \dots, \, g_m)
  = \left\{ (P_1, \, \dots, \, P_n) \in \mathbb{E}^n \,\vert\,
    g_i (P_1, \, \dots, \, P_n) = 0 \,\,\text{for all}\,\, 1 \leq i
    \leq m \right\} ,
  \label{eq:affine_variety}
\end{equation}
where $\mathbb{E}$ is the four-dimensional pseudo-Euclidean space,
$\mathbb{R}^{1, \, 3}$.

In missing energy events, not only in the double-sided decay topology,
the phase space can be decomposed into $\{(k_a$, $p_b)\}$.
By measuring the visible particle momenta $p_b$ while missing the
invisible particle momenta $k_a$, the phase space is projected onto
the visible subspace $\{p_b\}$.
The projection leads to {\em singularities} in the visible subspace,
where the full phase space $\{(k_a$, $p_b)\}$ is folded.
The event number density changes abruptly at the singularities, and they
can appear in several different forms such as wall or cusp.

The salient point of the algebraic singularity method is that the
singularities can be identified by constructing the Jacobian matrix of
the constraint polynomials $g_i$,
\begin{equation}
  J_{i j} = \frac{\partial g_i}{\partial k_j}.
  \label{eq:jacobian}
\end{equation}
Here, the index $j$ runs over the energy-momentum components of the
invisible particles, $k_a = (e_{a}$, $k_{ax}$, $k_{ay}$, $k_{az})$.
The algebraic singularity method is based on the observation that
the Jacobian matrix~\eqref{eq:jacobian} has {\em reduced rank} at
singularities.
In other words, one or more row vectors of the Jacobian matrix are
linearly dependent at singularities.
Note that the Jacobian matrix defines a linear mapping, which best
approximates $g_i$ at the point $k_a$.
If the Jacobian matrix has reduced rank, its image of the linear
mapping is smaller, and the approximation by the mapping is no
more valid.
Consequently, having reduced rank indicates a singular behavior of
$g_i$ near the point $k_a$.
In the subsequent works by different
authors~\cite{Rujula:2011qn, DeRujula:2012ns, Matchev:2019bon,
  Park:2020rol}, the reduced rank condition has been referred to as
``singularity condition.''
The Jacobian matrix is a square matrix if the number of polynomials in
Eq.~\eqref{eq:affine_variety} equals to that of unknowns.
In this case, the reduced rank condition is
equivalent to the vanishing determinant of the Jacobian matrix, as
studied in Refs.~\cite{Matchev:2019bon, Park:2020rol}.

In general, the defining polynomials of phase space form a system of
nonlinear coupled equations, and consequently, finding the reduced rank
condition of the Jacobian matrix is a nontrivial task.
Affine varieties are defined by ideals, which are the sets of all
polynomials having the same solution space.
The ideal generated by a finite set of polynomials is analogous to
the span of a finite number of vectors in linear algebra.
Indeed, the finite set of polynomials is termed a {\em basis} of the
ideal. A given ideal may have many different bases.
In Ref.~\cite{Kim:2009si},
it is emphasized that Gr\"obner bases, also known as the standard
bases, can be particularly useful for examining the reduced rank
condition of the Jacobian matrix because it can make the matrix a
row echelon form.
It is analogous to Gaussian elimination for linear systems.
Then, to identify the singularities, we set some of the diagonal
components of the Jacobian matrix in the Gr\"obner basis to be
vanishing and check whether the row vectors of the Jacobian matrix
with vanishing diagonal component are linearly dependent.
Unfortunately, the analytic calculation of Gr\"obner bases can be
challenging if the number of polynomials is large, and thus one
typically employs dedicated algorithms such as Buchberger's
algorithm~\cite{ref:Buchberger}.
In the case of the double-sided decay topology in~\eqref{eq:topology},
we can manage to obtain the corresponding Gr\"obner basis
analytically, as done in Ref.~\cite{Kim:2009si}. We will see the
expressions of the Gr\"obner basis in the following subsection.

Having identified the singularities by imposing the reduced rank
condition, we can construct an optimized kinematic variable, named
{\em singularity coordinate}.
It is required to be zero at singularities and is
perpendicular to the singularity hypersurface in the visible phase space.
Collider events near the singularities are projected onto the
coordinate. It is also normalized so that the events with the same
distance to the singularities give the same value.
It is called a {\em coordinate} as it determines the positions of
collider events on the singularity hypersurface.
Singularity coordinate is an implicit variable because it depends
on the masses of the unknown particles such as $Y$ and $\chi$
in the decay topology~\eqref{eq:topology}.
If trial mass values have correctly been chosen, the singularity
coordinate maximizes singular features.
Consequently, we can deduce the mass spectrum of unknown particles by
comparing several competing hypotheses using the singularity
coordinate.

Although the formal definition of singularity coordinates is given
in Ref.~\cite{Kim:2009si}, it lacks a practical recipe for
constructing them.
It has later been supplemented by the studies of applying the
algebraic singularity method to several physics processes of various
decay topologies: single $W$ production
$W \to \ell \nu$~\cite{Rujula:2011qn}, the dileptonic Higgs channel $h
\to W W \to \ell^+ \nu \ell^- \bar\nu$~\cite{DeRujula:2012ns}, heavy
Higgs boson decaying to a top-quark pair $H/A \to t \bar t \to bW \bar b
W \to b \ell^+ \nu \, \bar b \ell^- \bar \nu$~\cite{Park:2020rol},
and various single- and double-sided decay
topologies~\cite{Matchev:2019bon}.
In the studies, singularity coordinates are called interchangeably
with {\em singularity variables}.
The latter are also functions of visible particle momenta and unknown
particle masses, but it has mass dimension one.
The singularity variables enable us directly to measure the unknown
particle masses by identifying the peak or the endpoints of their
distributions.
By contrast, the singularity coordinates can have mass
dimensions much higher than one, and one can estimate the unknown
particle masses by performing template fitting of the distribution
shapes. We can hardly extract the mass information directly from the
distribution.
Therefore, singularity variables are more straightforward and
intuitive to use in physics analyses than singularity coordinates.
In this article, we distinguish the singularity coordinates and the
singularity variables and concentrate on the latter.

Before examining the singularity variable for the double-sided decay
topology in~\eqref{eq:topology}, we take a look into the singularity
variable for the simpler decay topology in order to help facilitate
the understanding of the algebraic singularity method.
Consider a single-sided two-body decay topology, where one of the
decay products is invisible,
\begin{equation}
  Y \longrightarrow v(p) \chi(k).
  \label{eq:topology_single}
\end{equation}
The kinematic constraints of the decay topology are given as follows:
\begin{align}
  k^2 &= M_\chi^2, \nonumber\\
  {(p + k)}^2 &= M_Y^2, \nonumber\\
  \vb*{k}_T &= \slashed{\vb*{P}}_T ,
\label{eq:single_two_body}
\end{align}
where $k = (e$, $\vb*{k}_T$, $k_z) = (e$, $k_x$, $k_y$, $k_z)$ contains
the unknowns of the system, and $p = (E$, $\vb*{p}_T$, $p_z) = (E$, $p_x$,
$p_y$, $p_z)$ is the visible particle momentum of the given collider event.
Using the lexicographic ordering $k_x \succ k_y \succ k_z \succ e$, we
find that the Gr\"obner basis of the system is given by
\begin{align}
    g_1 =&~ k_x - \slashed{P}_x, \nonumber\\
    g_2 =&~ k_y - \slashed{P}_y, \nonumber\\
    g_3 =& -2p_z k_z + 2 E e - \left( M_Y^2 - M_\chi^2 - m^2
      + 2 \vb*{p}_T \cdot \slashed{\vb*{P}}_T \right) , \nonumber\\
    g_4 =& - 4(E^2 - p_z^2) e^2 + 4 E \left( M_Y^2 - M_\chi^2 - m^2 + 2
      \vb*{p}_T \cdot \slashed{\vb*{P}}_T \right) e \nonumber\\
    & - (M_Y^2 - M_\chi^2 - m^2 + 2 \vb*{p}_T \cdot \slashed{\vb*{P}}_T )^2
      - 4 p_z^2 ( M_\chi^2 + \norm{\slashed{\vb*{P}}_T}^2 ) ,
\end{align}
where $m^2 = p^2$.
We have four constraint equations and four
unknowns. Therefore, the Jacobian is a $4 \times 4$ matrix given as follows:
\begin{equation}
  J_{ij} =
  \frac{\partial g_i}{\partial k_j} =
  \begin{pmatrix}
    1 & & & \\
    & 1 & & \\
    & & -2p_z & 2E \\
    & & & -8(E^2 - p_z^2) e + 4 E \left( M_Y^2 - M_\chi^2 - m^2 + 2
      \vb*{p}_T \cdot \slashed{\vb*{P}}_T \right)
  \end{pmatrix} . \label{eq:jac_single}
\end{equation}
As mentioned earlier, taking the Gr\"obner basis renders the Jacobian
matrix
having the row echelon form.
Except for the soft singularity where $p_z = 0$,
the reduced rank condition requires the fourth diagonal term of
$J$ to be vanishing,
\begin{equation}
  0 = J_{44} = 8 p_z E e \left( \frac{p_z}{E} - \frac{k_z}{e} \right)
  . \label{eq:single_j44}
\end{equation}
Here, we have used the constraints in~\eqref{eq:single_two_body} to
eliminate $M_Y^2$ and $\slashed{\vb*{P}}_T$.
Note that Eq.~\eqref{eq:single_j44} is equivalent to the relation that
\begin{equation}
  \frac{p_z}{E_T} = \frac{k_z}{e_T}
\end{equation}
if $p_z$, $E$, and $e$ are all nonzero. By using the relation, we get
\begin{align}
  M_Y^2
  = m^2 + M_\chi^2 + 2 \left( E_T e_T - \vb*{p}_T \cdot \vb*{k}_T
    \right) ,
\end{align}
where $\vb*{k}_T = \slashed{\vb*{P}}_T$.
The expression on the right-hand side is nothing but the transverse
mass squared for the $v$ and $\chi$ system.
It shows that the transverse mass corresponds to the singularity
variable of the decay topology~\eqref{eq:topology_single}, as shown in
Refs.~\cite{NewPhysicsWorkingGroup:2010hhy, Rujula:2011qn,
  DeRujula:2012ns, Matchev:2019bon}.
The transverse mass is the invariant mass in the $(1+2)$-dimensional
space, and the algebraic singularity method tells us that it is an
optimized variable. The Jacobian peak in the transverse mass
distribution is indeed a feature of singularity.

\subsection{Singularity variable for double-sided decay topology}

\noindent
After having recapitulated the $M_{T2}$ variable and the algebraic
singularity method, we now derive the singularity variable for
double-sided decay topology. The investigation of its relation with
the $M_{T2}$ variable is presented in the following section.
To derive the singularity variable using the algebraic singularity
method, we should first identify all the kinematic constraint
equations and the corresponding Gr\"obner basis.
For the double-sided decay topology
in~\eqref{eq:topology}, the kinematic constraints are given as
follows:\footnote{
  In the presence of multiple visible particles, the construction of
  kinematic constraints can be vulnerable to the combinatorial
  ambiguities on how to group the visible particles into separate
  sets.
  For instance, in the process of pair-produced gluinos decaying into
  jets, $\tilde g \tilde g \to j j \tilde\chi_1^0 + j j
  \tilde\chi_1^0$ with $j$ being a quark jet, there are three
  distinct ways of grouping the jets into two pairs of visible
  systems.
  We assume that the combinatorial ambiguity has been resolved by
  employing dedicated methods such as those proposed in
  Refs.~\cite{CMS:2007sch, Nojiri:2008hy, Rajaraman:2010hy,
    Baringer:2011nh, Choi:2011ys, Debnath:2017ktz}.
}
\begin{align}
    k_1^2 &= M_\chi^2, \nonumber\\
    k_2^2 &= M_\chi^2, \nonumber\\
    {(p_1 + k_1)}^2 &= M_Y^2, \nonumber\\
    {(p_2 + k_2)}^2 &= M_Y^2, \nonumber\\
  \vb*{k}_{1T} + \vb*{k}_{2T} &= \slashed{\vb*{P}}_T .
                                \label{eq:constraints_double_sided}
\end{align}
In this system, we have eight unknowns from the energy-momentum
components of invisible particle momenta $k_1$ and $k_2$, while there
are six constraints from~\eqref{eq:constraints_double_sided}.
Therefore, the system is underconstrained, and we
need additional constraints or ansatz to solve the system.
In the $M_{T2}$ variable,
the limitation has been overcome by the minimization over
$\vb*{k}_{1T}$, for a guessed value of $M_\chi$.
In the algebraic singularity method, we will see that the reduced rank
condition provides additional constraints to the unknowns.
Note that the number of additional constraints furnished by the
reduced rank condition depends on the form of diagonal components of
the Jacobian matrix.
As can be seen in the Jacobian matrix given in
Eq.~\eqref{eq:jac_single} and the matrix for the double-sided decay
topology, which will be shown shortly, not all diagonal components of
the Jacobian matrices can be set to be zero.
Unless we find a theorem concerning the relation among the leading terms
of Gr\"obner basis elements and decay topologies, it can be
deduced only by explicitly obtaining the Gr\"obner basis of the
polynomial equations of the given process.
Therefore, it is generally unclear whether the reduced rank condition
for a given decay topology would always provide additional constraints
enough to solve all the unknowns in conjunction with the known
kinematic constraints.
Nonetheless, we will see that we can obtain the constraints enough to
solve the unknowns for an input mass $M_\chi$ by using the reduced
rank condition in the case of the double-sided decay topology.

As we have done in Sec.~\ref{sec:mt2}, we eliminate $\vb*{k}_{2T}$ by
using the condition of missing transverse momentum. Then, we have
four constraints from the on-shell mass relations and six unknowns.
With the lexicographic ordering $e_1 \succ e_2 \succ k_{1z} \succ
k_{2z} \succ k_{1x} \succ k_{1y}$, the Gr\"obner basis is given by
\begin{align}
    g_1
    =&~ 2 E_1 e_1 - 2p_{1z} k_{1z} - 2 p_{1x} k_{1x} - 2 p_{1y} k_{1y}
    - (M_Y^2 - M_\chi^2 - m_1^2), \nonumber\\
    g_2
    =&~ 2 E_2 e_2 - 2p_{2z} k_{2z} + 2 p_{2x} k_{1x} + 2 p_{2y} k_{1y}
    - (M_Y^2 - M_\chi^2 - m_2^2 + 2 \vb*{p}_{2T} \cdot \slashed{\vb*{P}}_T), \nonumber\\
    g_3
    =& - 4 (E_{1}^2 - p_{1z}^2) k_{1z}^2 + 8 p_{1x} p_{1z} k_{1z}
       k_{1x} + 8 p_{1y} p_{1z} k_{1z} k_{1y} \nonumber\\
     & +4 (M_Y^2 - M_\chi^2 - m_1^2) p_{1z} k_{1z} \nonumber\\
    & - 4(E_{1}^2 - p_{1x}^2) k_{1x}^2 + 8 p_{1x} p_{1y} k_{1x} k_{1y}
      \nonumber\\
  & +4 (M_Y^2 - M_\chi^2 - m_1^2) p_{1x} k_{1x} \nonumber\\
     & - 4(E_{1}^2 - p_{1y}^2) k_{1y}^2
     +4 (M_Y^2 - M_\chi^2 - m_1^2) p_{1y} k_{1y} \nonumber\\
    & + (M_Y^2 - M_\chi^2 - m_1^2)^2 - 4 E_1^2 M_\chi^2 ,\nonumber\\
    g_4
    =& -4 (E_{2}^2 - p_{2z}^2) k_{2z}^2 - 8 p_{2x} p_{2z} k_{2z}
    k_{1x} - 8 p_{2y} p_{2z} k_{2z} k_{1y} \nonumber\\
    & + 4 (M_Y^2 - M_\chi^2 - m_2^2 + 2 \vb*{p}_{2T} \cdot
    \slashed{\vb*{P}}_T ) p_{2z} k_{2z} \nonumber\\
    & - 4 (E_{2}^2 - p_{2x}^2) k_{1x}^2 + 8 p_{2x} p_{2y} k_{1x}
    k_{1y} \nonumber\\
    & -4 \left[ (M_Y^2 - M_\chi^2 - m_2^2 + 2 \vb*{p}_{2T} \cdot
    \slashed{\vb*{P}}_T ) p_{2x} - 2 E_2^2 \slashed{P}_x \right]
  k_{1x} \nonumber\\
  & -4 (E_{2}^2 - p_{2y}^2) k_{1y}^2
  -4 \left[ (M_Y^2 - M_\chi^2 - m_2^2 + 2 \vb*{p}_{2T} \cdot
    \slashed{\vb*{P}}_T ) p_{2y} - 2 E_2^2 \slashed{P}_y \right]
  k_{1y} \nonumber\\
  & + (M_Y^2 - M_\chi^2 - m_2^2 + 2 \vb*{p}_{2T} \cdot
    \slashed{\vb*{P}}_T)^2 - 4 E_2^2 (M_\chi^2 + \norm{\slashed{\vb*{P}}_T}^2) .
\end{align}
Hypothesizing the values of $M_Y$ and $M_\chi$, we can solve the
polynomial equations to obtain the solution to the unknowns for a
given $\vb*{k}_{1T} = (k_{1x}$, $k_{1y})$.
The expressions of the solutions of $g_i$ are given as follows:
\begin{align}
    e_a
    &= \frac{A_a + p_{az} k_{az}}{E_a} , \nonumber\\
    k_{az}
    &= \frac{p_{az} A_a \pm E_a \sqrt{A_a^2 - E_{aT}^2
        e_{aT}^2}}{E_{aT}^2},
      \label{eq:onshell_sols}
\end{align}
and $\vb*{k}_{2T} = \slashed{\vb*{P}}_T - \vb*{k}_{1T}$. Here,
\begin{equation}
  A_a = \frac{M_Y^2 - M_\chi^2 - m_a^2}{2} + \vb*{p}_{aT} \cdot
  \vb*{k}_{aT}
  \label{eq:Afunc}
\end{equation}
is the same as that defined in Ref.~\cite{Cho:2008tj}.
Due to the quadratic equations $g_3$ and $g_4$, we have two degenerate
solutions for each longitudinal component.

The Jacobian matrix of the Gr\"obner basis is the $4 \times 6$ matrix
given by
\begin{equation}
  J =
  \begin{pmatrix}
    2 E_1 & & -2 p_{1z} & & -2 p_{1x} & -2 p_{1y} \\
    & 2 E_2 & & - 2p_{2z} & 2 p_{2x} & 2 p_{2y} \\
    & & \partial g_3 / \partial k_{1z} &
    & \partial g_3 / \partial k_{1x}
    & \partial g_3 / \partial k_{1y} \\
    & & & \partial g_4 / \partial k_{2z}
    & \partial g_4 / \partial k_{1x}
    & \partial g_4 / \partial k_{1y}
  \end{pmatrix} ,
  \label{eq:jacobian_double}
\end{equation}
where
\begin{align}
    \frac{\partial g_3}{\partial k_{1z}}
  &= 8 p_{1z} (A_1 + p_{1z} k_{1z} ) - 8 E_1^2 k_{1z}, \nonumber\\
    \frac{\partial g_3}{\partial k_{1x}}
  &= 8 p_{1x} (A_1 + p_{1z} k_{1z} ) - 8 E_1^2 k_{1x} ,
    \nonumber\\
    \frac{\partial g_3}{\partial k_{1y}}
  &= 8 p_{1y} (A_1 + p_{1z} k_{1z} ) - 8 E_1^2 k_{1y} ,
    \nonumber\\
    \frac{\partial g_4}{\partial k_{2z}}
  &= 8 p_{2z} (A_2 + p_{2z} k_{2z}) - 8 E_2^2 k_{2z}, \nonumber\\
  \frac{\partial g_4}{\partial k_{1x}}
  &= - 8 p_{2x} (A_2 + p_{2z} k_{2z}) + 8 E_2^2 (\slashed{P}_x -
    k_{1x}) ,
    \nonumber\\
  \frac{\partial g_4}{\partial k_{1y}}
  &= - 8 p_{2y} (A_2 + p_{2z} k_{2z}) + 8 E_2^2 (\slashed{P}_y -
    k_{1y}) .
\end{align}
Except for the soft singularities where $E_1 = 0$ and $E_2 = 0$,
the reduced rank condition for the Jacobian matrix in
Eq.~\eqref{eq:jacobian_double} implies that
\begin{align}
  \frac{\partial g_3}{\partial k_{1z}}
  &= 0, \label{eq:diag_1} \\
  \frac{\partial g_4}{\partial k_{2z}}
  &= 0, \label{eq:diag_2} \\
  \det
  \begin{pmatrix}
    \partial g_3 / \partial k_{1x} & \partial g_3 / \partial k_{1y} \\
    \partial g_4 / \partial k_{1x} & \partial g_4 / \partial k_{1y}
  \end{pmatrix}
  &= 0 . \label{eq:singularity_det}
\end{align}
Note that from the vanishing diagonal terms, i.e., Eqs.~\eqref{eq:diag_1}
and~\eqref{eq:diag_2}, each longitudinal component is uniquely
determined as
\begin{equation}
  k_{az} = \frac{p_{az} A_a}{E_{aT}^2} .
  \label{eq:sol_kz}
\end{equation}
The expressions of the longitudinal momenta are the same as in the
modified MAOS method proposed in Ref.~\cite{Park:2011uz}.
By comparing the longitudinal components with those in
Eq.~\eqref{eq:onshell_sols}, we find that the condition of vanishing
diagonal terms is equivalent to
\begin{equation}
  A_a = E_{aT} e_{aT} .
  \label{eq:A_singularity}
\end{equation}
From the definition of $A_a$ given in Eq.~\eqref{eq:Afunc},
the first two of the above conditions turn out to be
\begin{align}
  M_Y^2 &= m_1^2 + M_\chi^2 + 2 (E_{1T} e_{1T} - \vb*{p}_{1T} \cdot
          \vb*{k}_{1T}), \nonumber\\
  M_Y^2 &= m_2^2 + M_\chi^2 + 2 (E_{2T} e_{2T} - \vb*{p}_{2T} \cdot
          \vb*{k}_{2T}),
          \label{eq:MY_singularity}
\end{align}
which are the transverse masses of $v_1 \chi$ and $v_2 \bar\chi$
systems.
Interestingly, we have arrived at a similar conclusion as in the
previous discussion: the transverse masses are the singularity
variables of the double-sided decay topology where there exists an
invisible particle in each decay chain.
Because we here consider symmetric decay chains, the
condition~\eqref{eq:MY_singularity} corresponds to the balanced
configuration, where
\begin{equation}
  M_{1T} = M_{2T} .
  \label{eq:singularity_condition_1}
\end{equation}
Because $\vb*{k}_{2T} = \slashed{\vb*{P}}_{T} - \vb*{k}_{1T}$, this
condition is an equation of two variables, $k_{1x}$ and $k_{1y}$.

The condition~\eqref{eq:singularity_det} is a new one that does not
appear in the derivation of the singularity variable for
single-sided decay topology.
After eliminating $M_Y^2$ using $A_1$ in $\partial g_3 / \partial
\vb*{k}_{1T}$ and using $A_2$ in $\partial g_4 / \partial
\vb*{k}_{1T}$, then substituting the singularity solutions into the
invisible longitudinal momenta given in~\eqref{eq:sol_kz}, we get
\begin{multline}
  \det
  \begin{pmatrix}
    \partial g_3 / \partial k_{1x} & \partial g_3 / \partial k_{1y} \\
    \partial g_4 / \partial k_{1x} & \partial g_4 / \partial k_{1y}
  \end{pmatrix} \\
  = \frac{64 E_1^2 E_2^2}{E_{1T}^2 E_{2T}^2} \det
  \begin{pmatrix}
    A_1 p_{1x} - E_{1T}^2 k_{1x}
    & A_1 p_{1y} - E_{1T}^2 k_{1y} \\
    - A_2 p_{2x} + E_{2T}^2 (\slashed{P}_x - k_{1x})
    & - A_2 p_{2y} + E_{2T}^2 (\slashed{P}_y - k_{1y})
  \end{pmatrix} .
\end{multline}
Then, by using the condition in~\eqref{eq:A_singularity}, we arrive at
\begin{align}
  0 &= \det
  \begin{pmatrix}
    \partial g_3 / \partial k_{1x} & \partial g_3 / \partial k_{1y} \\
    \partial g_4 / \partial k_{1x} & \partial g_4 / \partial k_{1y}
  \end{pmatrix} \nonumber\\
  &=
  \frac{16 E_1^2 E_2^2 e_{1T} e_{2T}}{E_{1T} E_{2T}} \det
  \begin{pmatrix}
    \partial M_{1T}^2 / \partial k_{1x}
    & \partial M_{1T}^2 / \partial k_{1y} \\
    \partial M_{2T}^2 / \partial k_{1x}
    & \partial M_{2T}^2 / \partial k_{1y}
  \end{pmatrix} , \label{eq:singularity_condition_2}
\end{align}
where we have used the expressions of the gradients of the transverse
masses given in~\eqref{eq:MT_gradient}.
If all the transverse energies $E_{aT}$ and $e_{aT}$ are nonzero, the
reduced rank condition implies that the determinant of the Jacobian
matrix for the transverse masses is vanishing at singularities.

Summarizing our findings so far, the reduced rank condition for the
double-sided decay topology has led to two conditions at singularity
points:
\begin{enumerate}[(i)]
\item the balanced configuration of the transverse masses,
  Eq.~\eqref{eq:singularity_condition_1},
\item and the vanishing determinant of the Jacobian matrix for the
  transverse masses, Eq.~\eqref{eq:singularity_condition_2}.
\end{enumerate}
Because the conditions provide two equations, we can solve
them to obtain the unknowns, $k_{1x}$ and $k_{1y}$, for trial mass
$M_\chi$.\footnote{
  For asymmetric decay chains, that is, $M_{Y} \neq M_{\bar
    Y}$, the reduced rank condition gives us three equations, two from
  the transverse masses and one from the vanishing determinant of the
  Jacobian matrix. The system is underconstrained unless we have ansatz
  for either $M_Y$ or $M_{\bar Y}$.
}
This is in contrast with the $M_{T2}$ variable, which is obtained by
the minimization of the objective function~\eqref{eq:MT2_objective}
over $k_{1x}$ and $k_{1y}$.

\section{\label{sec:comp}Comparison of \boldmath{$M_{T2}$} and
  singularity variables}

\noindent
We are now in a position to examine the singularity variable of
double-sided decay topology by comparing it with the $M_{T2}$
variable.
As we have seen in Sec.~\ref{sec:mt2},
the convexity of the $M_{T2}$ objective function ensures that the
global minimum of the objective function and the corresponding
solutions to the transverse momenta of the invisible particles are
uniquely determined for each event.
On the other hand, Eqs.~\eqref{eq:singularity_condition_1}
and~\eqref{eq:singularity_condition_2} are a coupled nonlinear system,
which may yield multiple solutions to the invisible particle
momenta.
As a result of the minimization,
the endpoint of the $M_{T2}$ distribution is smaller than or equal to the
parent particle, as in Eq.~\eqref{eq:MT2_endpoint}.
The reduced rank condition, however, does not warrant such a feature
on the singularity variables.
Moreover, it is challenging to solve
Eqs.~\eqref{eq:singularity_condition_1} and
~\eqref{eq:singularity_condition_2} analytically by transforming or
reducing it to a simpler form.
It is not even clear to see if the solution of the equations would be
unique.
Therefore, we rely on sequential numerical methods to obtain
the solutions of the equations.

Because both $M_{T2}$ and singularity variables do not have analytic
expressions in general cases, the comparison of the two different
types of variables is not a straightforward task.
If either or both of $\partial M_{1T} / \partial \vb*{k}_{1T}$ and
$\partial M_{2T} / \partial \vb*{k}_{1T}$ are vanishing,
the condition~\eqref{eq:singularity_condition_2} is trivially
satisfied,
but it does not always lead to the balanced configuration, $M_{1T} =
M_{2T}$.
In the case where visible particle systems are massless and upstream
momentum is vanishing, we have seen that the $M_{T2}$ solution to the
invisible particle momenta in balanced configurations is given by
Eq.~\eqref{eq:mt2_ksol}.
Indeed, when $m_1 = m_2 = 0$, the condition~\eqref{eq:mt2_bal_cond}
always holds, and thus every collider event is in a balanced
configuration.
One can see that the analytic expression of the $M_{T2}$
solution~\eqref{eq:mt2_ksol} satisfies
both conditions, Eqs.~\eqref{eq:singularity_condition_1}
and~\eqref{eq:singularity_condition_2}.
Therefore, in this particular case, we have proven that
the singularity variables {\em contain} the $M_{T2}$ variable.

\begin{figure}[tb!]
  \begin{center}
    \includegraphics[width=0.46\textwidth]{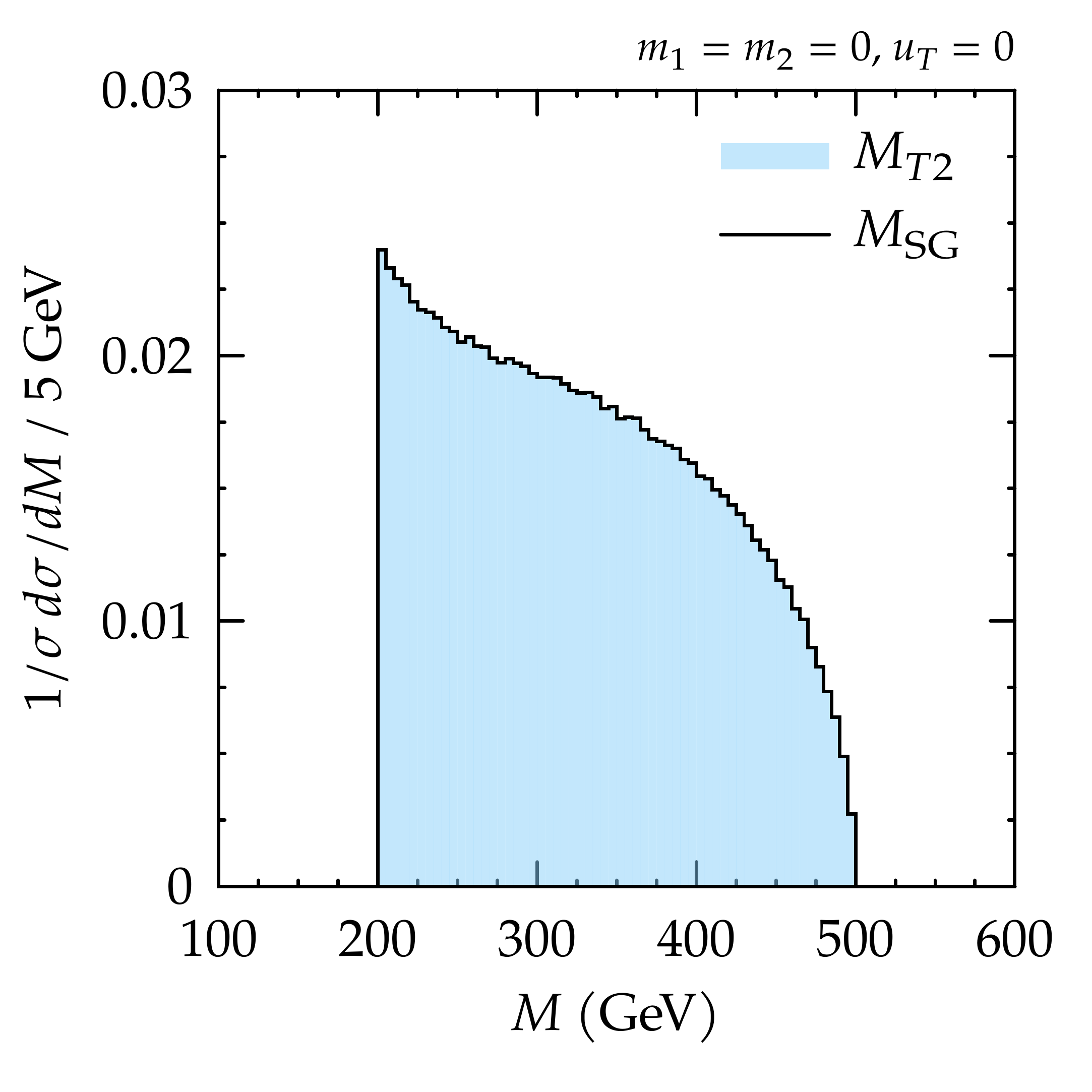}
    \includegraphics[width=0.46\textwidth]{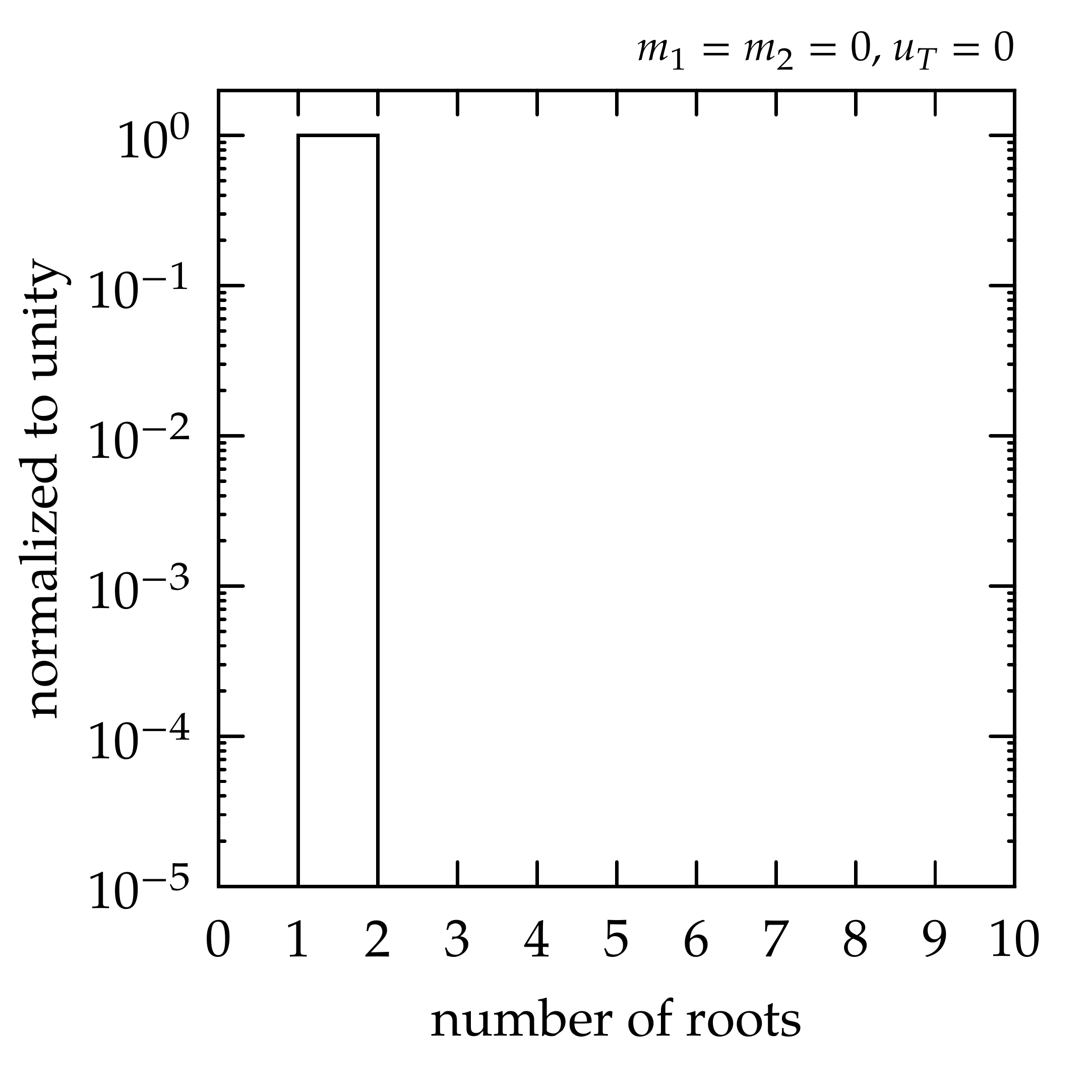}\\
    \includegraphics[width=0.46\textwidth]{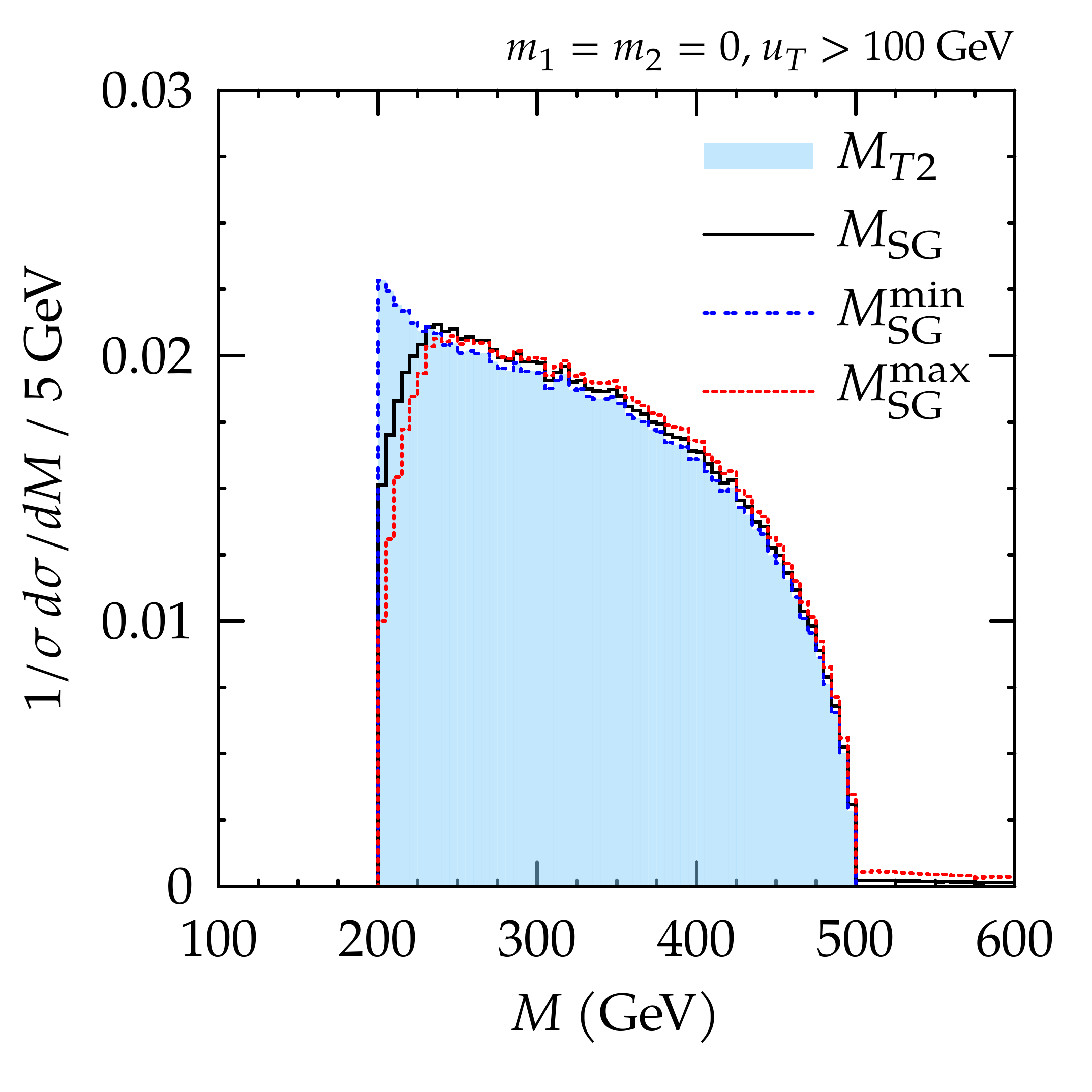}
    \includegraphics[width=0.46\textwidth]{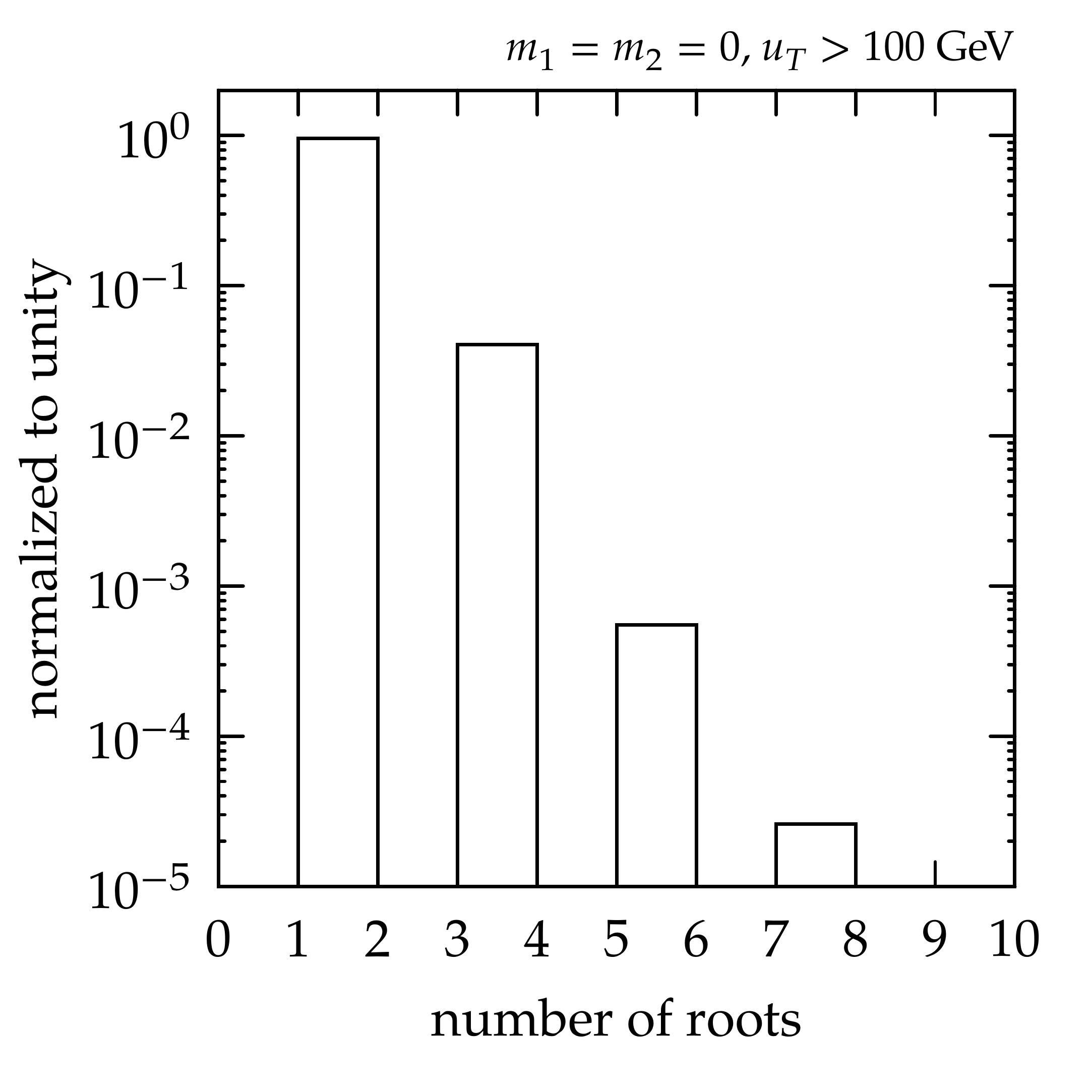}
  \end{center}
  \caption{\label{fig:two_body}
    Distributions of the singularity variables $M_\text{SG}$ and
    $M_{T2}$ (left) and the number of roots of reduced rank conditions
    (right) for phase-space events of $M_Y = 500$~GeV and $M_\chi =
    200$~GeV.
    In the upper panels, the upstream momentum $u_T$ is vanishing,
    while in the lower panels, it is larger than 100~GeV.
    The visible particle systems are all massless, $m_1 = m_2 = 0$,
    and we have taken $M_\chi = M_\chi^\text{true}$.
  }
\end{figure}

For comparing the two types of collider variables, we have
generated one million phase-space event sample and calculated the
variables by employing numerical algorithms.
As our study concentrates on the pure comparison of the $M_{T2}$ and
singularity variables, we do not take into account realistic detector
effects, such as particle momentum smearing and misidentification
rates, in our simulation.
For calculating $M_{T2}$, we used the public
software package based on the bisection method~\cite{Lester:2014yga}.
As we are not aware of any software package for singularity
variables, we have calculated them by using our own code
implementation.
As noted earlier, Eq.~\eqref{eq:singularity_condition_2} is not the
polynomial equation of $\vb*{k}_{1T}$, and it can possess multiple
real or complex solutions.
To solve Eqs.~\eqref{eq:singularity_condition_1}
and~\eqref{eq:singularity_condition_2},
we repeat Newton's method with a large
number of heuristically chosen initial guesses for $\vb*{k}_{1T}$ in
our code implementation.
For a cross-check, we have compared the results from our code with
those from the \texttt{NSolve} method of
\texttt{Mathematica}\textsuperscript{TM}, which can find multiple roots of
nonlinear equations.
We will release our code implementation for calculating the
singularity variables as a public software package in a separate
publication.

In Fig.~\ref{fig:two_body}, we display the distributions of the
singularity and the $M_{T2}$ variables for phase-space events of $M_Y =
500$~GeV and $M_\chi = 200$~GeV. The visible particle systems are
taken to be massless.
In the figure, $M_\text{SG}$ denotes the
singularity variables in~\eqref{eq:MY_singularity}, i.e.,
\begin{equation}
  M_\text{SG} = M_{1T} = M_{2T} = \frac{M_{1T} + M_{2T}}{2} .
\end{equation}
When there exist multiple solutions for invisible particle momenta in
an event, we will have multiple values of $M_\text{SG}$.
In this case, we can take the minimum and the maximum among the
$M_\text{SG}$ values into separate consideration,
\begin{equation}
 M_\text{SG}^\mathrm{min} \leq M_\text{SG} \leq M_\text{SG}^\mathrm{max} .
\end{equation}
If $M_\text{SG}$ were invariant masses, the parent particle mass $M_Y$
would be bounded as $M_\text{SG}^\mathrm{min} \leq M_Y \leq
M_\text{SG}^\mathrm{max}$, as in the case of the antler decay topology
studied in Refs.~\cite{Matchev:2019bon, Park:2020rol}.
However, the mass hierarchy does not hold because both
$M_\text{SG}^\mathrm{min}$ and $M_\text{SG}^\mathrm{max}$ could be
smaller than $M_Y$ for $M_\text{SG}$ being transverse masses.
In the upper panels of Fig.~\ref{fig:two_body}, we show the
distributions in the case of vanishing upstream momentum.
As already noted, in this case, the $M_{T2}$ solution to the invisible
particle momenta satisfies the reduced rank condition of the algebraic
singularity method.
Thus, one might expect that one of the $M_\text{SG}$ values would be
in accord with the $M_{T2}$ value while the others would have
different values.
Our numerical study shows that the singularity solution to the
invisible particle momenta is unique for all events and $M_{T2} =
M_\text{SG}$.
We conjecture that this observation led the author of
Ref.~\cite{Kim:2009si} to a conclusion: ``the $M_{T2}$ variable could
be derived from the algebraic singularity method.''
However, to arrive at the right conclusion, we should also investigate
the more general cases with nonvanishing upstream momentum and nonzero
visible particle masses.
The distributions for events with large upstream momentum $u_T >
100$~GeV are shown in the lower panels of Fig.~\ref{fig:two_body}.
In this case, we find that there exist events with multiple solutions,
although a vast majority of events possess unique solutions to the
reduced rank condition.
In the $M_\text{SG}$ distribution, we have appropriately weighted the
histogram for the events having multiple solutions.
We observe that even if multiple solutions exist, the minimum
among the $M_\text{SG}$ values is still identical to the $M_{T2}$
value, i.e.,
\begin{equation}
  M_\text{SG} \geq M_{T2}
  \label{eq:SG_T2_relation}
\end{equation}
for all events.
It implies that some of the $M_\text{SG}$ values can exceed the true
parent particle mass $M_Y^\text{true}$ because the $M_{T2}$ distribution is
bounded from above by $M_Y^\text{true}$ for $M_\chi = M_\chi^\text{true}$.
However, the rate of events with $M_\text{SG} > M_Y^\text{true}$
is quite small, as can be seen in the lower left panel of
Fig.~\ref{fig:two_body}.

\begin{figure}[tb!]
  \begin{center}
    \includegraphics[width=0.46\textwidth]{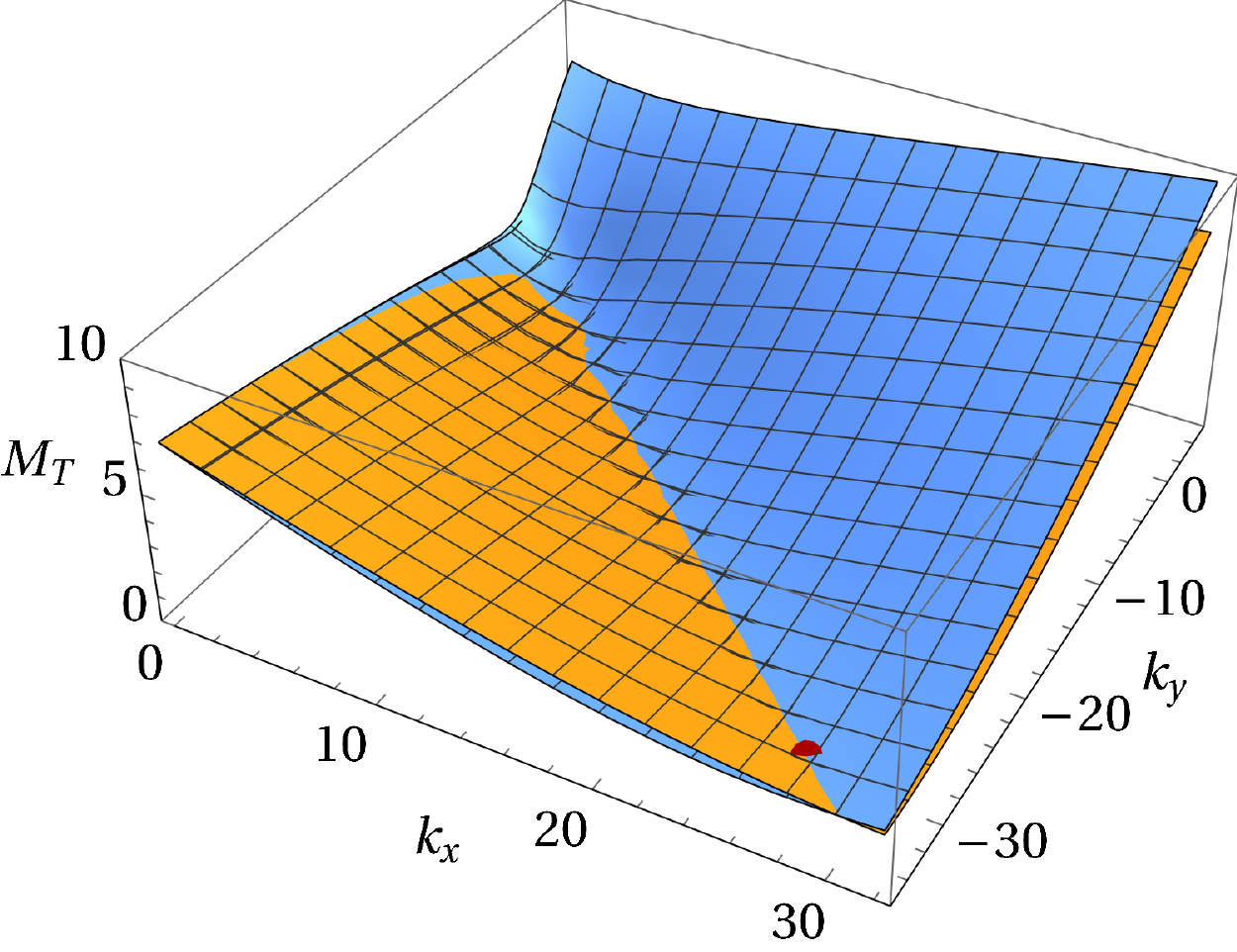}
    \includegraphics[width=0.46\textwidth]{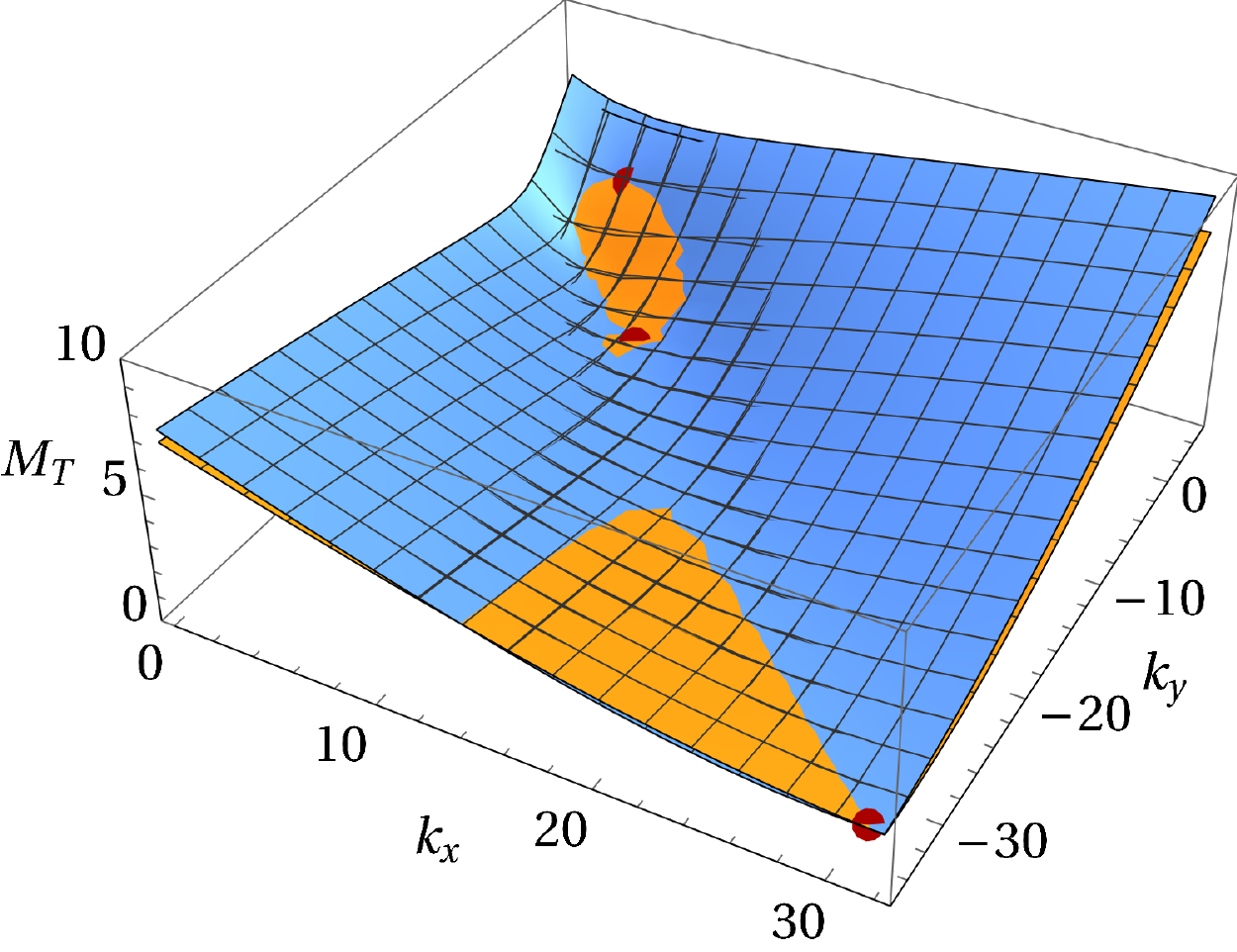}
  \end{center}
  \caption{\label{fig:MT_2D}
    Transverse masses $M_{1T}$ and $M_{2T}$ for an event with
    vanishing (left) and nonvanishing upstream momentum (right).
    The red points in each plot correspond to the solutions to the
    reduced rank condition of the algebraic singularity method.
    The scales of axes are arbitrarily normalized.
  }
\end{figure}

We lack the understanding of the condition for having multiple
solutions to the reduced rank condition unless we find the analytic
expressions of the solutions to Eqs.~\eqref{eq:singularity_condition_1}
and~\eqref{eq:singularity_condition_2}.
Still, we may acquire a rough understanding of the multiple solutions
by looking at the shapes of the transverse mass functions in the
presence of upstream momentum.
In Fig.~\ref{fig:MT_2D}, we show the transverse masses in the space of
$(k_{1x}$, $k_{1y})$ for an event with vanishing or nonvanishing
upstream momentum.
In the case where upstream momentum is vanishing, the points of
balanced configurations $M_{1T} = M_{2T}$ lie on a connected curve.
Among the points on the curve, the iterations of the $M_{T2}$
calculation will stop at a point that gives the minimum of the objective
function~\eqref{eq:MT2_objective}.
In the singularity method, we alternatively look for points satisfying
the vanishing determinant
condition~\eqref{eq:singularity_condition_2}.
We have found that the two different methods pick out an identical
point when upstream momentum is vanishing. See the left panel of
Fig.~\ref{fig:MT_2D}.
On the other hand, if upstream momentum is nonvanishing, the curves of
balanced configuration can appear in disconnected regions in the space
of invisible particle momentum.
The $M_{T2}$ variable will still take one point among them, where the
objective function is minimized, whereas several points on different
curves can satisfy the reduced rank condition, as can be seen in the
right panel of Fig.~\ref{fig:MT_2D}.
In this situation, we will have multiple solutions in the algebraic
singularity method.
Our numerical study shows that one of the multiple solutions is still
identical to the point picked out by the $M_{T2}$ variable in balanced
configurations.
We note that the physical interpretation of the multiple solutions to
the reduced rank condition is still unknown, and we are in eager
pursuit of new insights into the better understanding of the multiple
solutions.

\begin{figure}[tb!]
  \begin{center}
    \includegraphics[height=0.46\textwidth]{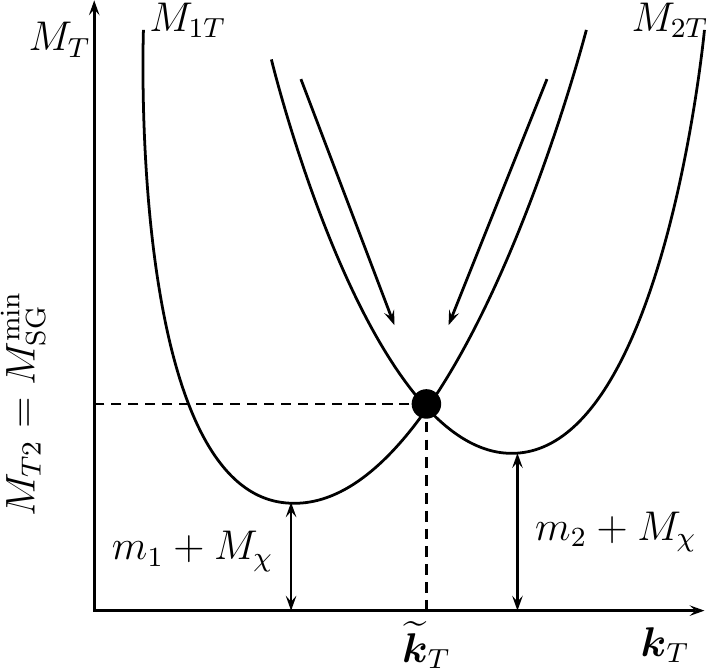}
    \includegraphics[height=0.46\textwidth]{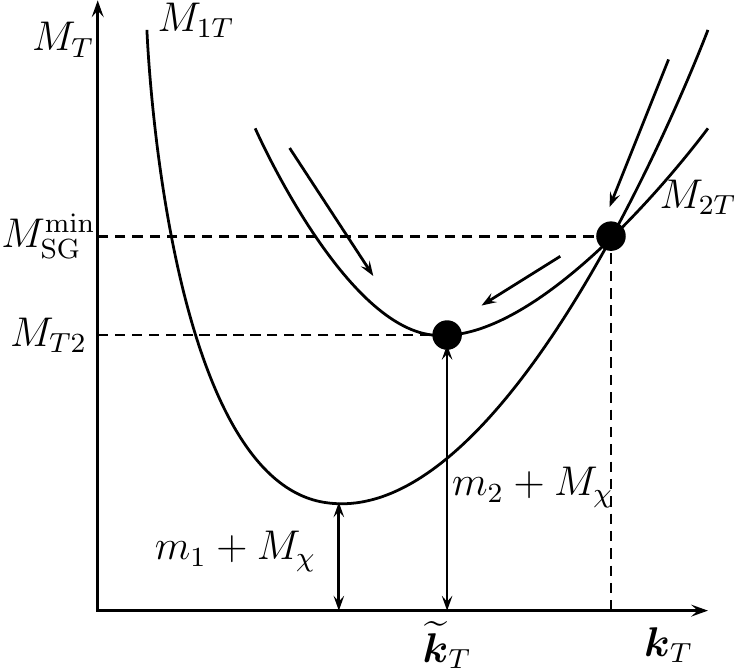}
  \end{center}
  \caption{\label{fig:msg_conf}
    The same pictures as in Fig.~\ref{fig:mt2_conf}, but with
    depicting the minimum value of the singularity variable
    $M_\text{SG}^\mathrm{min}$.
    The black dots in each plot correspond to either or both of the
    $M_{T2}$ and the $M_\text{SG}^\mathrm{min}$ values.
    $\widetilde{\vb*{k}}_T$ denotes the $M_{T2}$ solution to the
    invisible particle momenta.
  }
\end{figure}

We now turn our attention to the case of massive visible particle
systems. We have generated separate phase-space event sample while
keeping the $M_Y$ and $M_\chi$ values unchanged.
Now, the masses of the visible particle system, $m_1$ and $m_2$, are
set to be varying between $0$ and $M_Y - M_\chi$ as in the
three-body decay process.
A notable difference in comparison to the massless case is the
presence of unbalanced configurations.
Recall that unbalanced configurations where $M_{1T} \neq M_{2T}$ occur
when the condition~\eqref{eq:mt2_bal_cond} does not hold.
About 35\% of our event sample is in unbalanced configurations.
In such events, the $M_{T2}$ solution can never satisfy the reduced
rank condition because it violates the condition in
Eq.~\eqref{eq:singularity_condition_1}.
On the other hand, in the case of balanced configuration, we can
expect the same relation as in the massless case.
The locations of $M_\text{SG}^\mathrm{min}$ in two different
configurations are depicted in Fig.~\ref{fig:msg_conf}.
The right panel of Fig.~\ref{fig:msg_conf} shows that
$M_\text{SG}^\mathrm{min}$ can be larger than the $M_{T2}$ value in
unbalanced configurations.
There is another case the figure does not show: the singularity
variable goes {\em undefined} if the collider event is in an
unbalanced configuration and the transverse masses never meet at any
point in the physical domain of $\vb*{k}_{1T}$.
Even in this case, the $M_{T2}$ value is given by either of the
stationary points of $M_{1T}$ and $M_{2T}$.

\begin{figure}[tb!]
  \begin{center}
    \includegraphics[width=0.46\textwidth]{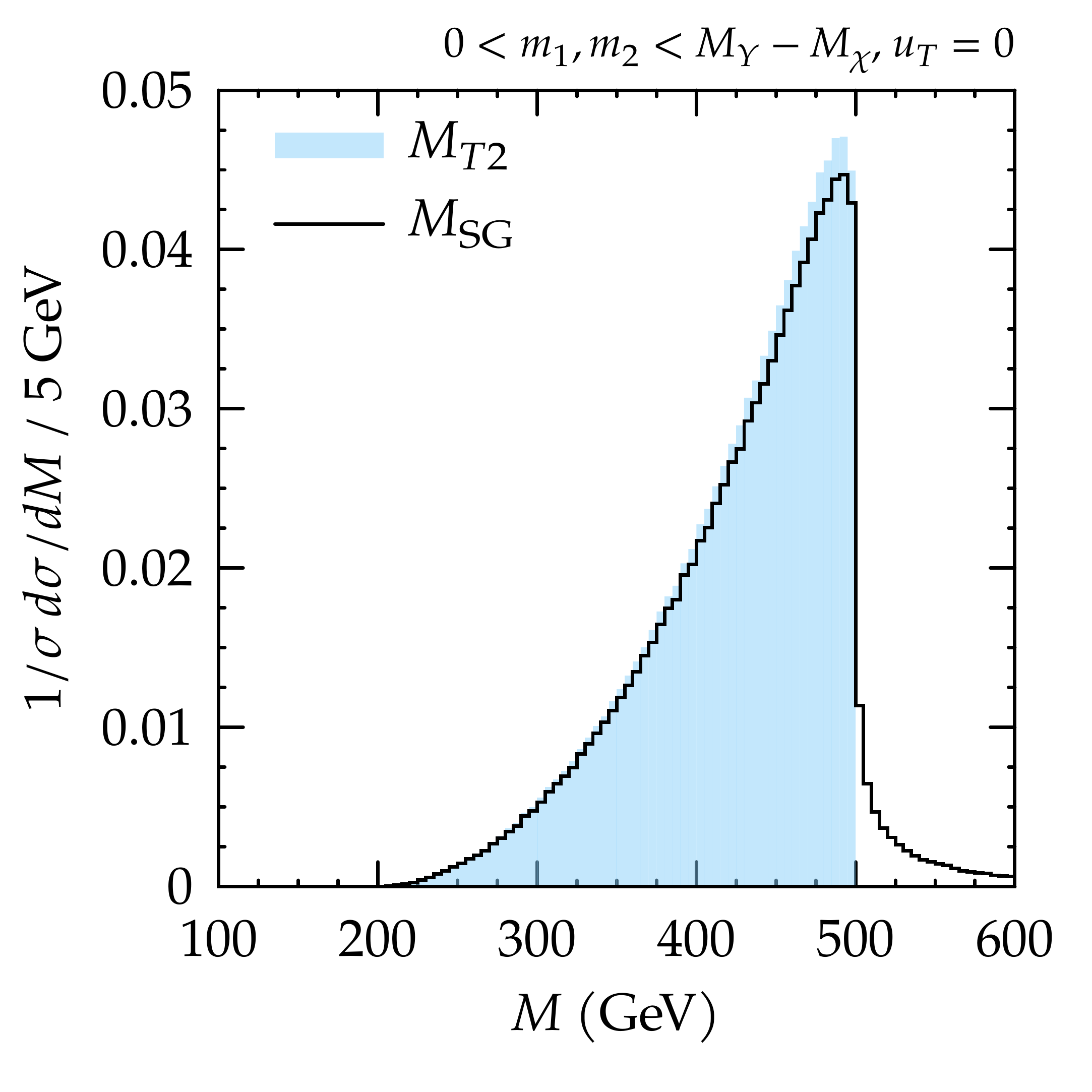}
    \includegraphics[width=0.46\textwidth]{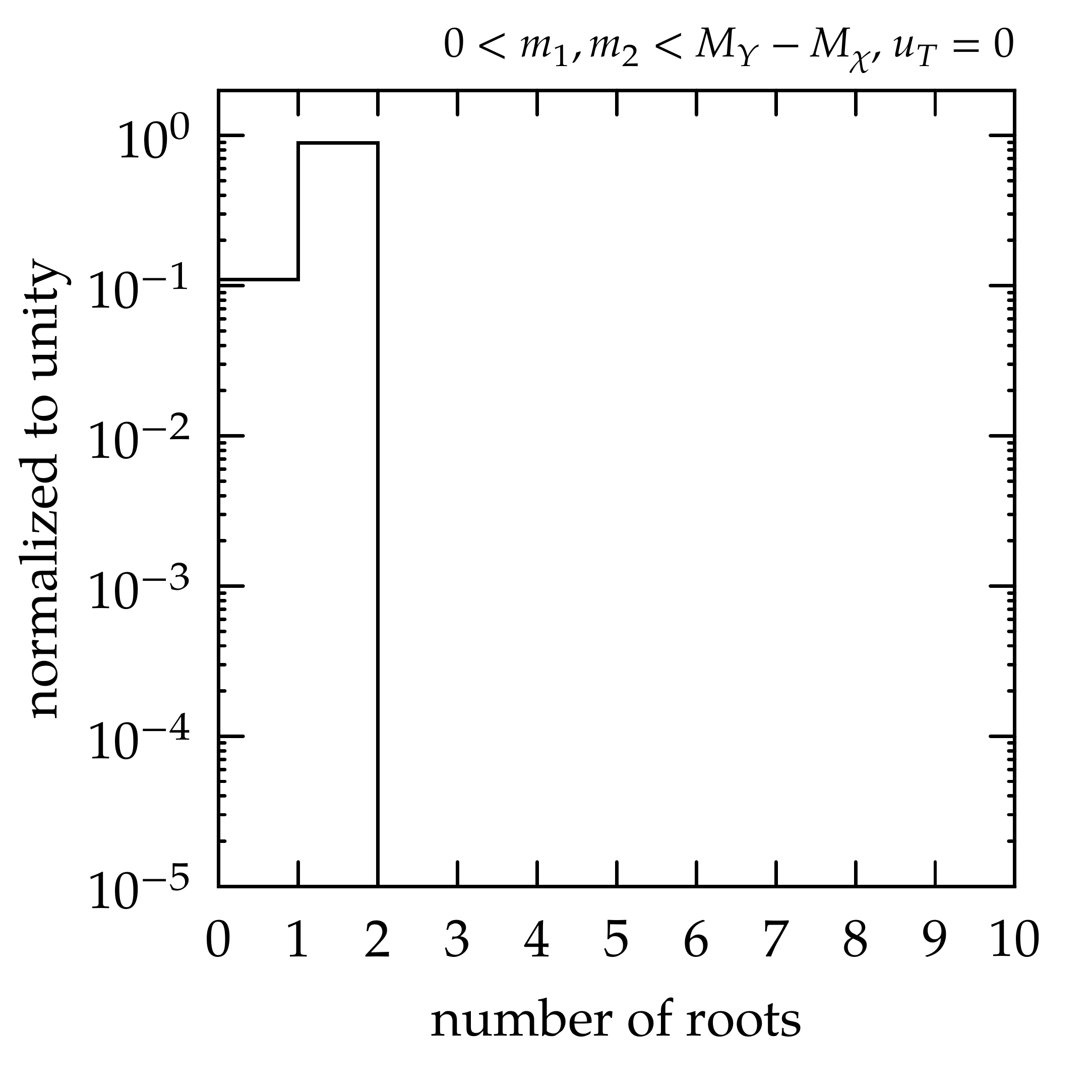}\\
    \includegraphics[width=0.46\textwidth]{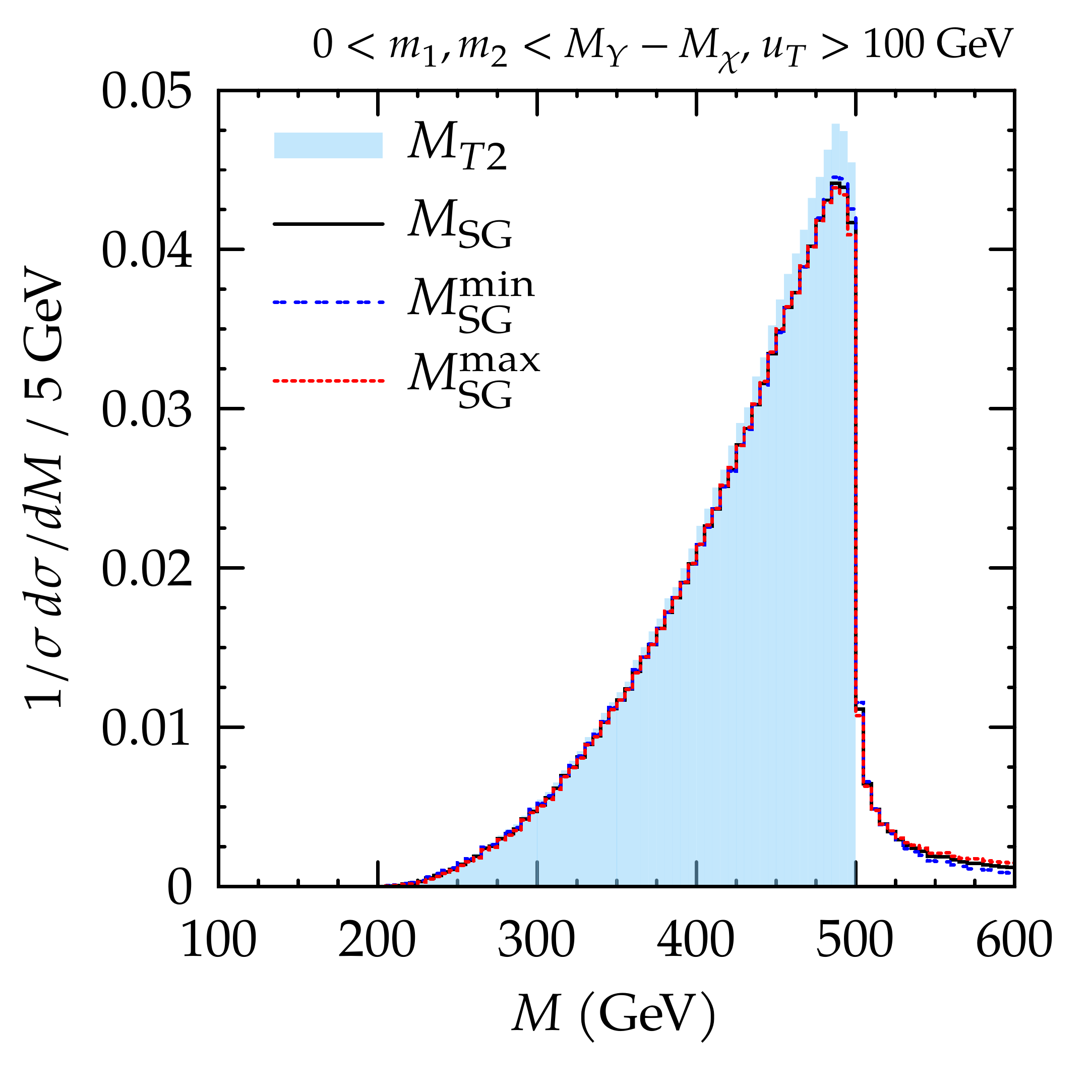}
    \includegraphics[width=0.46\textwidth]{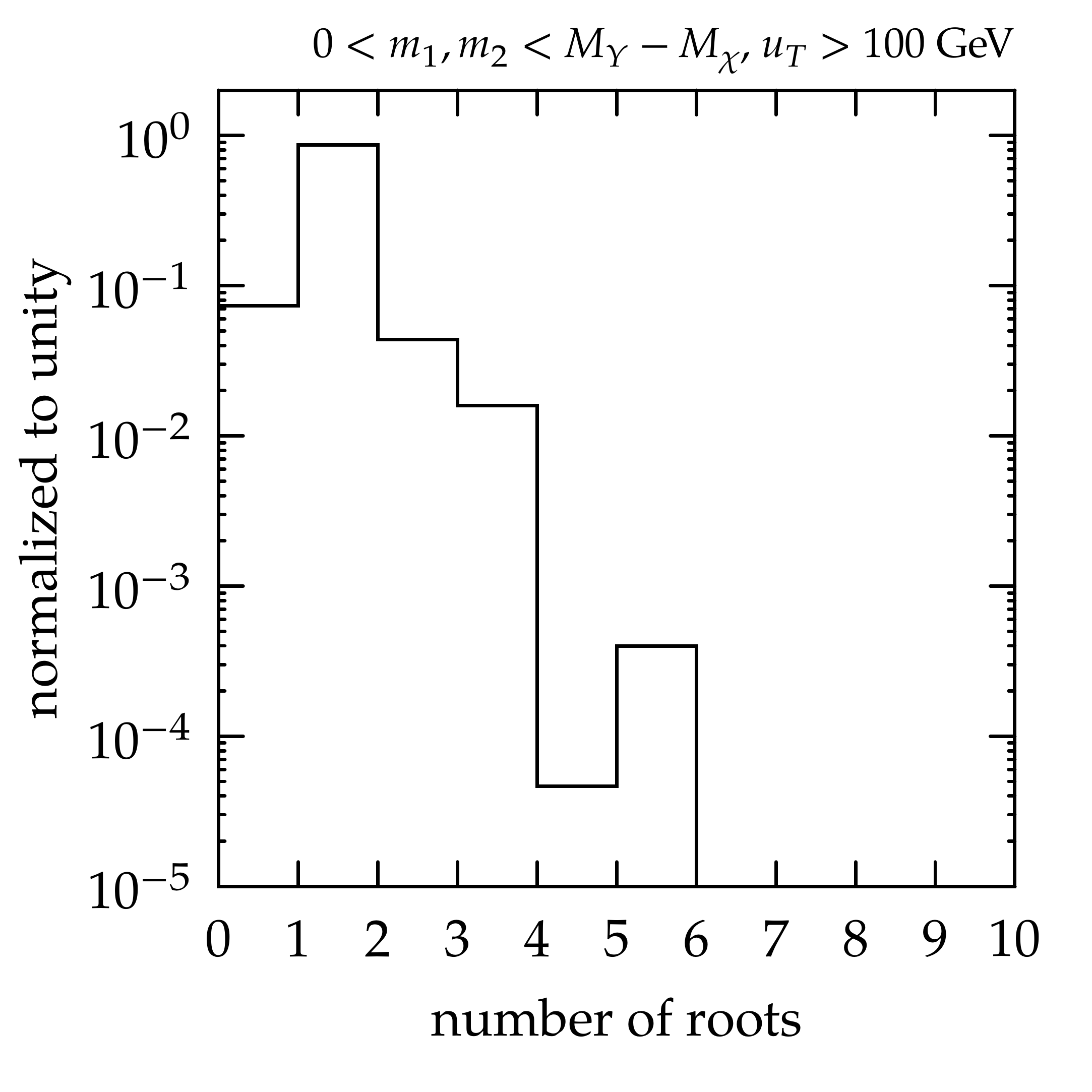}
  \end{center}
  \caption{\label{fig:three_body}
    Distributions of the singularity variables $M_\text{SG}$ and
    $M_{T2}$ (left) and the number of roots of reduced rank conditions
    (right) for phase-space events of $M_Y = 500$~GeV and $M_\chi =
    200$~GeV.
    The masses of visible particle systems are varying between
    $0$ and $M_Y - M_\chi$, and we have taken $M_\chi =
    M_\chi^\text{true}$.
    In the upper panels, the upstream momentum $u_T$ is vanishing,
    while in the lower panels, it is larger than 100~GeV.
  }
\end{figure}

The $M_{T2}$ and the $M_\text{SG}$ distributions for phase-space event
data with massive visible particles are shown in
Fig.~\ref{fig:three_body}.
We find that the singularity variable is undefined for about 10\% of
events.\footnote{
  One way of resolving the problem of undefined singularity variable is
  to relax the condition of balanced
  configuration~\eqref{eq:singularity_condition_1}.
  We may use the minimum of $\abs{M_{1T} - M_{2T}}$ (or ${(M_{1T} -
  M_{2T})}^2$) subject to the condition of vanishing determinant
  in~\eqref{eq:singularity_condition_2}.
  We have implemented the modified singularity variable in our
  numerical study and found it is identical to $M_{T2}$.
}
For comparison, we show the distributions for the events
where the singularity variable can be attained.
As in the case of massless visible particle systems, the singularity
variable is uniquely determined when upstream momentum is vanishing.
However, due to the presence of unbalanced configurations, the
endpoint of the $M_\text{SG}$ distribution exceeds the parent particle
mass $M_Y$, as can be seen in the upper left panel of
Fig.~\ref{fig:three_body}.
On the other hand, the reduced rank condition can yield multiple
solutions in the case where upstream momentum is nonvanishing, even
though most of the events still possess unique solutions.
In both cases, we observe that the $M_\text{SG}$ distributions have a
sharp peak around the parent particle mass. Therefore, we expect that
the singularity variable can be useful for the discovery or the mass
measurement of new particles at collider experiments.

\subsection{Beyond trivial zero}

\noindent
We finally consider an interesting case where the visible and the
invisible particles are all massless, $m_1 = m_2 = M_\chi = 0$.
We often deal with the fully massless case in both Standard Model (SM)
measurements and new physics searches.
For example, in the $W W \to \ell^+ \nu \ell^- \bar \nu$ process,
which we will investigate in this subsection, all the final-state
particles are massless.
In the searches for supersymmetric processes, e.g., pair produced
sleptons decaying to charged leptons and the lightest neutralinos, the
final-state particles are often assumed to be massless in the
simplified model approach.
See, for instance, Refs.~\cite{ATLAS:2019lff, CMS:2020bfa}.

In Sec.~\ref{sec:mt2}, we have seen that the $M_{T2}$ solution to
invisible particle momenta would simply be given by
Eq.~\eqref{eq:mt2_ksol_massless} when upstream momentum is vanishing.
In the case of nonzero upstream momentum, the missing transverse
momentum $\slashed{\vb*{P}}_T$ may lie inside the smaller sector
bounded by visible particle momenta $\vb*{p}_{1T}$ and $\vb*{p}_{2T}$.
In this situation, the $M_{T2}$ value is zero because the invisible
momenta is taken to be $\vb*{k}_{aT} \propto \vb*{p}_{aT}$.
This is called a {\em trivial zero} of $M_{T2}$~\cite{Lester:2011nj}.
It is not a coincidence but an unexpected consequence of minimization
inherent to the definition of $M_{T2}$.

\begin{figure}[tb!]
  \begin{center}
    \includegraphics[height=0.46\textwidth]{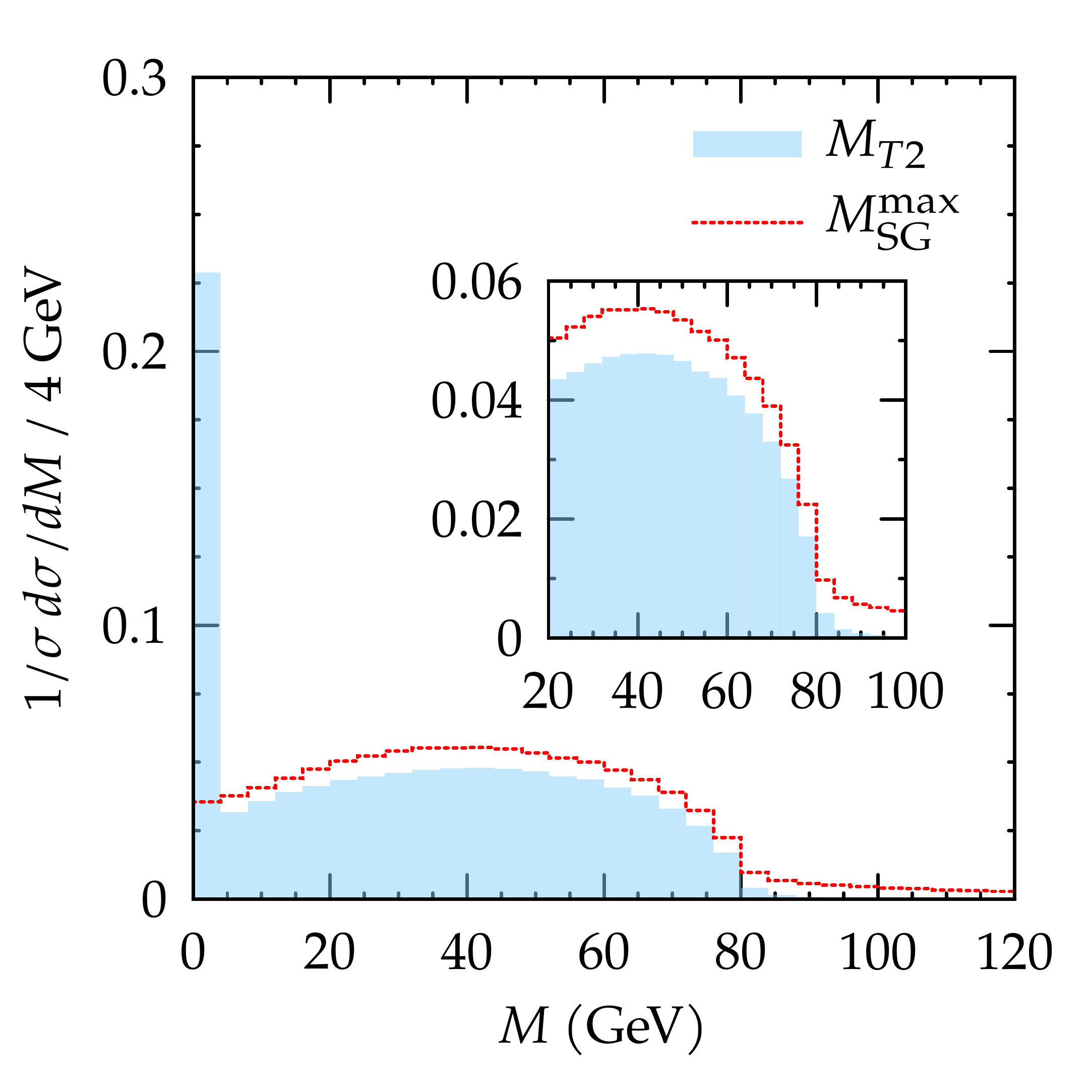}
  \end{center}
  \caption{\label{fig:ww_ttbar}
    Distributions of $M_{T2}$ and $M_\text{SG}^\mathrm{max}$ in the
    $WW$ subsystem of top pair process. The insert in the plot shows
    the distributions in the mass region between $20$ and $100$~GeV.
  }
\end{figure}

To investigate the trivial zero of $M_{T2}$, we have generated
Monte Carlo event samples of the dileptonic top pair process
by using \texttt{Pythia}~\cite{Sjostrand:2014zea}.
The proton-proton collision energy has been set to be 13~TeV.
We take the $WW$ subsystem of the top-pair decay events, so the
upstream momentum is given by $\vb*{u}_T = \vb*{p}_T^b +
\vb*{p}_T^{\bar b}$ plus the transverse momentum of initial state
radiations.
The distributions of $M_{T2}$ and $M_\text{SG}^\mathrm{max}$ are shown
in Fig.~\ref{fig:ww_ttbar}.
For the $WW$ subsystem, every event is in a balanced configuration.
Therefore, $M_\text{SG}^\mathrm{min} = M_{T2}$ for all events.
The $M_{T2}$ distribution has a peak around $M_{T2} = 0$, which
is due to the trivial zero of $M_{T2}$.
Meanwhile, the $M_\text{SG}^\mathrm{max}$ distribution does not have a
peak at zero, and it has a smoothly-falling shape around the $W$ boson
mass.
Consequently, the $M_\text{SG}^\mathrm{max}$ distribution has a
more number of events around the $W$ boson mass than the $M_{T2}$
distribution, as can be seen in the insert plot in
Fig.~\ref{fig:ww_ttbar}.

From this observation, we expect that using the
$M_\text{SG}^\mathrm{max}$ distribution in the fully massless case
could be more advantageous than $M_{T2}$ in the situation where the
trivial zero of $M_{T2}$ affects the statistical significance of new
physics searches.
For instance, if one takes the lightest neutralino mass to be zero,
many signal events of the slepton pair process will have the $M_{T2}$
value at zero, and one of the dominant backgrounds, the SM $WW +$ jet
process, will also have a high population in the same bin.
It can be circumvented by using $M_\text{SG}^\mathrm{max}$, which is
free from trivial zero, and its distribution has an edge around the
hypothesized slepton mass.

\section{Conclusions}

\noindent
The $M_{T2}$ variable is served as the observable of choice
when analyzing the collider events of double-sided decay topology with
invisible particles in the final state.
Meanwhile, the algebraic singularity method is still less studied
compared to the $M_{T2}$ variable.
It is a framework for studying any decay topology containing
invisible particles by exploiting all the available kinematic
constraints.
We can derive singularity variables from the reduced rank condition of
the algebraic singularity method. The singularity variables are
supposed to be optimal because they can capture singular features
characteristic to the decay topology under consideration.

The question investigated in this article is whether the $M_{T2}$
variable is also a singularity variable, or we can derive another
kinematic variables more optimal than the $M_{T2}$ by applying the
algebraic singularity method to the double-sided decay topology.
The analytic comparison of the $M_{T2}$ variable and the
corresponding singularity variable is challenging because their
analytic expressions are not available yet.
However, in some special cases, e.g., when the visible particle
systems are massless and upstream momentum is vanishing, we have found
that the $M_{T2}$ variable is in accord with the singularity variable
$M_\text{SG}$.
In the other cases, multiple singularity solutions can exist, and we
demonstrated that the $M_{T2}$ variable is still contained in the
singularity variables in many cases.
Indeed, it corresponds to the minimum of the
singularity variables, $M_{T2} \leq M_\text{SG}$, except for the
unbalanced configurations of $M_{T2}$.
In the latter case, the $M_{T2}$ value can be distinct from the
singularity variables.
We hope that this article would serve as supplementary material
for studying the algebraic singularity method.

The singularity variables can also be compared with another type of
kinematic variables, $M_2$, which is an extension of $M_{T2}$ to the
$(1+3)$-dimensional space~\cite{Cho:2014naa}.
The singularity coordinate has been derived in
Ref.~\cite{Matchev:2019bon}, but detailed analysis using the Gr\"obner
basis is missing.
We will perform the analysis in our future publication.

Another aspect that has not been considered in this article is the
effects of a realistic collider environment, such as particle momentum
resolutions, misidentification rates, and combinatorial uncertainties
in the presence of multiple visible particles.
They should be properly examined for practical applications of the
algebraic singularity method to physics analyses at colliders.
We stress that the study deserves follow-up works for concrete physics
processes, including realistic simulations with backgrounds.

\section*{Acknowledgments}

The author is grateful to Seodong~Shin for valuable comments on the
manuscript.
This work was supported by Institute for Basic Science (IBS) under the
project code, IBS-R018-D1.

\bibliography{msg}

\end{document}